\documentclass[a4paper,fleqn,usenatbib]{mnras}

\usepackage{newtxtext,newtxmath}
\usepackage[T1]{fontenc}
\usepackage{ae,aecompl}

\usepackage{graphicx}	% Including figure files
\usepackage{amsmath}	% Advanced maths commands
\usepackage{amssymb}	% Extra maths symbols

\newcommand{\eg}{e.g., }
\newcommand{\ie}{i.e., }
\newcommand{\Msun}{M_{\odot}}
\newcommand{\kms}{km~s$^{-1}$}

\newcommand{\Mej}{M_{\rm ej}}

\newcommand{\Ye}{Y_{\rm e}}
\def\gsim{\mathrel{\rlap{\lower 4pt \hbox{\hskip 1pt $\sim$}}\raise 1pt \hbox {$>$}}}
\def\lsim{\mathrel{\rlap{\lower 4pt \hbox{\hskip 1pt $\sim$}}\raise 1pt \hbox {$<$}}}

\def\ion#1#2{{\rm #1}~{\sc #2}}

\title[Systematic Opacity Calculations for Kilonovae]{Systematic Opacity Calculations for Kilonovae}

\author[M. Tanaka et al.]{
Masaomi Tanaka,$^{1}$\thanks{E-mail: masaomi.tanaka@astr.tohoku.ac.jp}
Daiji Kato,$^{2,3}$
Gediminas Gaigalas,$^{4}$
Kyohei Kawaguchi$^{5}$
\\
$^{1}$Astronomical Institute, Tohoku University, Sendai 980-8578, Japan\\
$^{2}$National Institute for Fusion Science, 322-6 Oroshi-cho, Toki 509-5292, Japan\\
$^{3}$Department of Advanced Energy Engineering Science, Kyushu University, Kasuga, Fukuoka 816-8580, Japan\\
$^{4}$Institute of Theoretical Physics and Astronomy, Vilnius University, Saul\.{e}tekio Ave. 3, Vilnius, Lithuania\\
$^{5}$Institute for Cosmic Ray Research, The University of Tokyo, 5-1-5 Kashiwanoha, Kashiwa, Chiba 277-8582, Japan
}

\begin{document}
\label{firstpage}
\pagerange{\pageref{firstpage}--\pageref{lastpage}}
\maketitle

\begin{abstract}
  Coalescence of neutron stars gives rise to kilonova, thermal emission
  powered by radioactive decays of freshly synthesized $r$-process nuclei.
  Although observational properties are largely affected by bound-bound
  opacities of $r$-process elements, available atomic data have been limited.
  In this paper, we study element-to-element variation of the opacities
    in the ejecta of neutron star mergers
    by performing systematic atomic structure calculations
    of $r$-process elements for the first time.
  We show that the distributions of energy levels
  tend to be higher as electron occupation increases for each electron shell
  due to the larger energy spacing
  caused by larger effects of spin-orbit and electron-electron interactions.
  As a result, elements with a fewer number of electrons in the outermost shells
  tend to give larger contributions to the bound-bound opacities.
  This implies that Fe is not representative for the opacities
    of light $r$-process elements.
  The average opacities for the mixture of $r$-process elements
 are found to be 
    $\kappa \sim 20-30 \ {\rm cm^2 \ g^{-1}}$ for the electron fraction of $\Ye \le 0.20$,
  $\kappa \sim 3-5  \ {\rm cm^2 \ g^{-1}}$ for $\Ye = 0.25-0.35$,
  and $\kappa \sim 1  \ {\rm cm^2 \ g^{-1}}$ for $\Ye = 0.40$
  at $T = 5,000-10,000$ K, and they steeply decrease at lower temperature.
  We show that, even with the same abundance or $\Ye$,
    the opacity in the ejecta changes with time
    by one order of magnitude from 1 to 10 days after the merger.
    Our radiative transfer simulations with the new opacity data
    confirm that ejecta with a high electron fraction
    ($\Ye \gsim 0.25$, with no lanthanide)
    are needed to explain the early, blue emission in GW170817/AT2017gfo
    while lanthanide-rich ejecta
    (with a mass fraction of lanthanides $\sim 5 \times 10^{-3}$)
    reproduce the long-lasting near-infrared emission.
\end{abstract}


%%%%%%%%%%%%%%%%%%%%%%%%%%%%%%%%%%%%%%%%%%%%%%%%%%

%%%%%%%%%%%%%%%%% BODY OF PAPER %%%%%%%%%%%%%%%%%%

\section{Introduction}
\label{sec:intro}

Coalescence of neutron stars (NSs) is a phenomenon of interest
in a wide area in astrophysics: it is one of the primary
targets of gravitational wave (GW) observations,
a candidate progenitor of short gamma-ray bursts (GRBs),
and a possible origin of the $r$-process elements in the Universe.
In fact, the detection of gravitational waves from a NS merger
has been achieved for the first time in 2017 (GW170817, \citealt{abbott17}).
Subsequent electromagnetic (EM) observations over
a wide wavelength range \citep{abbott17MMA}
identified the counterpart AT2017gfo, and 
provided rich information including the link between NS mergers and GRBs
\citep{abbott17GRB} and $r$-process nucleosynthesis by the NS merger.

In particular, intensive observations have been performed for
AT2017gfo in the ultraviolet, optical, and infrared wavelengths
\citep[\eg][]{andreoni17,arcavi17,chornock17,coulter17,cowperthwaite17,diaz17,drout17,evans17,kasliwal17,kilpatrick17,lipunov17,muccully17,nicholl17,pian17,shappee17,siebert17,smartt17,soares-santos17,tanvir17,tominaga18,troja17,utsumi17,valenti17}.
The observed properties are broadly consistent with kilonova
\citep[\eg]{kasen17,tanaka17,perego17,rosswog18},
thermal emission powered by radioactive decays of newly synthesized
$r$-process elements (\citealt{li98,kulkarni05,metzger10},
see \citealt{rosswog15,tanaka16,fernandez16,metzger17} for reviews).
The presence of the ``red'' (near-infrared, NIR) component
implies that the ejecta are composed of lanthanide elements
\citep{kasen13,tanaka13}.
On the other hand, the ``blue'' (optical) component suggests that
lighter $r$-process elements are also synthesized
\citep{metzger14,kasen15,tanaka18}.
These multiple ejecta components are naturally expected
in numerical simulations
\citep[see e.g.,][]{shibata17,perego17,kawaguchi18,radice18a,shibata19}.

Although $r$-process nucleosynthesis is confirmed in GW170817/AT2017gfo,
the exact abundance pattern synthesized by the NS merger is not yet clear.
The most straightforward ways are identifying elements
in the observed spectra and measuring their abundances.
However, due to the large Doppler shift,
blend of many absorption lines, and incompleteness of the atomic data,
identification of the all the observed spectral features are
  challenging
\citep[see][for the identification of Sr II lines]{watson19}.
%conclusive identification is not yet done \citep[see \eg][]{kasen17,smartt17}.

Another method is modelling the light curves.
For a simple one-zone model with a constant opacity,
the typical peak time of the light curve 
scales as $t_{\rm peak} \propto \kappa^{0.5}$,
where $\kappa$ is the opacity.
Accordingly, the peak luminosity scales as $L_{\rm peak} \propto \kappa^{-0.65}$
if the radioactive decay luminosity decreases with $t^{-1.3}$,
which is typical for neutron star mergers where
$\beta$ decays of many $r$-process nuclei are involved
\citep[\eg][]{metzger10,hotokezaka17heat}.
Since $\kappa$ is sensitive to the element abundances (see below),
we can indirectly infer the abundance from the light curves.
In fact, many attempts of light curve modelling
have been performed by assuming simple,
constant opacities \citep[\eg][]{cowperthwaite17,villar17,perego17}.
However, the opacities in the NS merger ejecta heavily depend
on the wavelengths, and evolve with time by reflecting
the changes in density, temperature, and thus, ionization/excitation states.
Therefore, to connect the abundance patterns in the ejecta
with the observed properties,
we need to consider detailed atomic opacities of $r$-process elements

In fact, understanding of atomic opacities of $r$-process elements
in kilonova has grown in the past several years.
\citet{kasen13} first performed atomic structure calculations for
selected lanthanide elements while
\citet{tanaka13} compiled available data for $r$-process elements.
  They found that, as in the case of Fe-rich ejecta of Type Ia supernovae \citep{pinto00},
  the main contribution of the opacities come from bound-bound transitions
  of heavy elements, \ie
  bound-free, free-free, and electron scattering opacitieis are subdominant.
They also found high bound-bound opacities of lanthanide elements,
which make kilonova fainter and redder than previously expected.
Then, atomic structure calculations for selected lanthanide elements
and lighter $r$-process elements have been performed by 
\citet{fontes17}, \citet{wollaeger18}, and \citet{tanaka18}.
More recently, \citet{kasen17} and \citet{fontes20}
provided atomic calculations of all the lanthanide elements.

However, available atomic calculations still do not cover
  many $r$-process elements that NS mergers synthesize.
  Due to this situation, opacities of some representative elements have been
  used to compensate lacking data in detailed radiative transfer simulations.
  But the previous studies showed that, for example, the opacities of
  Nd and Er are different although they are both lanthanide elements
  \citep{tanaka18,fontes20}.
  Because of the lack of systematic atomic calculations,
  it has not been clear how large element-to-element variations
  of the opacities exist across the wide range of $r$-process elements,
  and how these variations affect the opacities of the NS merger ejecta.

In this paper, we perform the systematic opacity calculations
  of the $r$-process elements to
  understand the elemental variation of the opacities and physics behind it.
  As in the previous studies of kilonova opacity
  \citep[\eg][]{kasen13,fontes17,tanaka18},
  providing very accurate atomic data (so-called spectroscopic accuracies,
  which are required for the opacities of stellar interior
  or stellar atmosphere), is beyond our scope
  since it is not yet computationally feasible
  \citep[see][for efforts on selected elements]{gaigalas19,radziute20}.  
  On the other hand, we aim at providing the {\it complete} dataset
  for the bound-bound opacities of $r$-process elements.
  In Section \ref{sec:atomic}, we show results of atomic structure calculations.
We discuss elemental dependence of the opacities in Section \ref{sec:opacity}.
Then, we apply the opacity data for radiative transfer simulations
in Section \ref{sec:kilonova}.
Finally we give a summary in Section \ref{sec:summary}.
Throughout of the paper, magnitudes are given in AB magnitude system.

\section{Atomic calculations}
\label{sec:atomic}

\subsection{Methods}

We perform systematic atomic calculations for the elements
from Fe ($Z = 26$) to Ra ($Z = 88$).
In this paper, we mainly focus on kilonova emission at
$t \gsim 1$ day after the merger
(hereafter $t$ denotes time after the merger).
In such timescale, the temperature in the ejecta
is $T \lsim 20,000 $ K, at which typical ionization stages
of heavy elements are either neutral,
or singly to triply ionized states (I - IV).
We calculate atomic energy levels and radiative transitions
of these ions using HULLAC (Hebrew University Lawrence Livermore Atomic Code,
\citealt{bar-shalom01}).

Since the calculation methods are same as in \citet{tanaka18},
we give only a brief overview of the calculations.
In the HULLAC code, the orbital functions are derived
by solving the single electron Dirac equation
with a central-field potential which includes
both a nuclear field and a spherically averaged
potential due to electron-electron interactions.
Then, $N$-electron configuration state functions are
constructed by coupled anti-symmetric products of the orbital functions.
Relativistic configuration interaction (RCI) calculations are performed
with the configuration state functions.
In the RCI calculations, we include the ground-state
and low excited-state configurations,
which have a bulk contribution to the bound-bound opacity.
The configurations used in our calculations are summarized in Table \ref{tab:hullac}.
The total Hamiltonian consisting of the Dirac-Coulomb Hamitonian,
the Breit interaction and the leading QED corretions is diagonalized with
the multi configuration state functions, and atomic energy levels
are obtained as eigenvalues of the total Hamiltonian.
Electric-dipole transition probabilities are calculated in length (Babushkin) gauge.
  
In the HULLAC code, the central-field potential is constructed
from an electron charge distribution of the Slater-type orbital
(see Equation (3) of \citealt{tanaka18}),
for which we use the ground state configuration of the next higher charge state.
The potential is optimized so that
the first-order configuration average energies of the ground state
and low-lying excited states are minimized.
The configurations used for the energy minimization are shown in bold in Table \ref{tab:hullac}.
To perform systematic calculations,
we normally choose only the ground configuration for the energy minimization.
However, we also include other configurations
when the lowest energy for each configuration
significantly deviates from that in the NIST Atomic Spectra Database
(ASD, \citealt{kramida18}).

%%%%%%%%%%%%%%%%%%%%%%%%%%%%%%%%%%%%%%%%%%%%%%%%%%%%
% Figure
%%%%%%%%%%%%%%%%%%%%%%%%%%%%%%%%%%%%%%%%%%%%%%%%%%%%
\begin{figure}
  \begin{center}
    \includegraphics[scale=0.75]{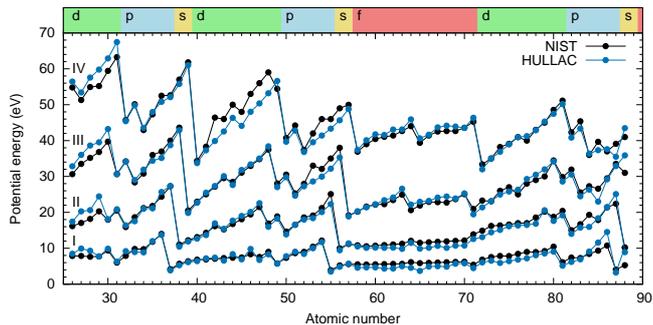}
\caption{
  \label{fig:epot}
  Comparison between the calculated ionization potentials
  using the HULLAC code
  (blue) and the those given in the NIST ASD (black, \citealt{kramida18}).
  }
\end{center}
\end{figure}
%%%%%%%%%%%%%%%%%%%%%%%%%%%%%%%%%%%%%%%%%%%%%%%%%%%%

%%%%%%%%%%%%%%%%%%%%%%%%%%%%%%%%%%%%%%%%%%%%%%%%%%%%
% Figure
%%%%%%%%%%%%%%%%%%%%%%%%%%%%%%%%%%%%%%%%%%%%%%%%%%%%
\begin{figure}
  \begin{center}
    \includegraphics[scale=0.75]{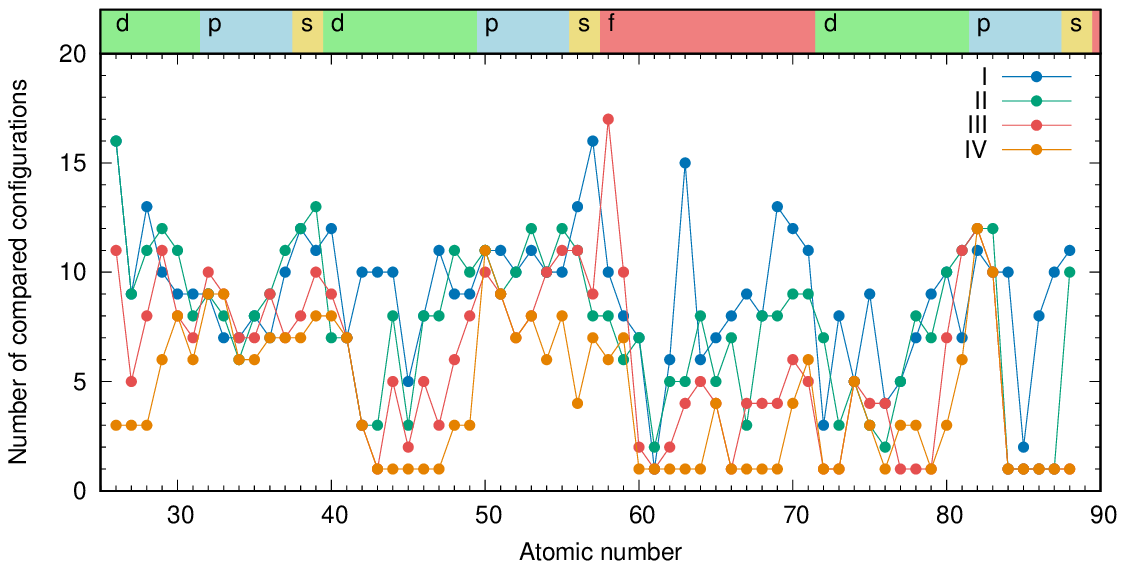}\\
    \includegraphics[scale=0.75]{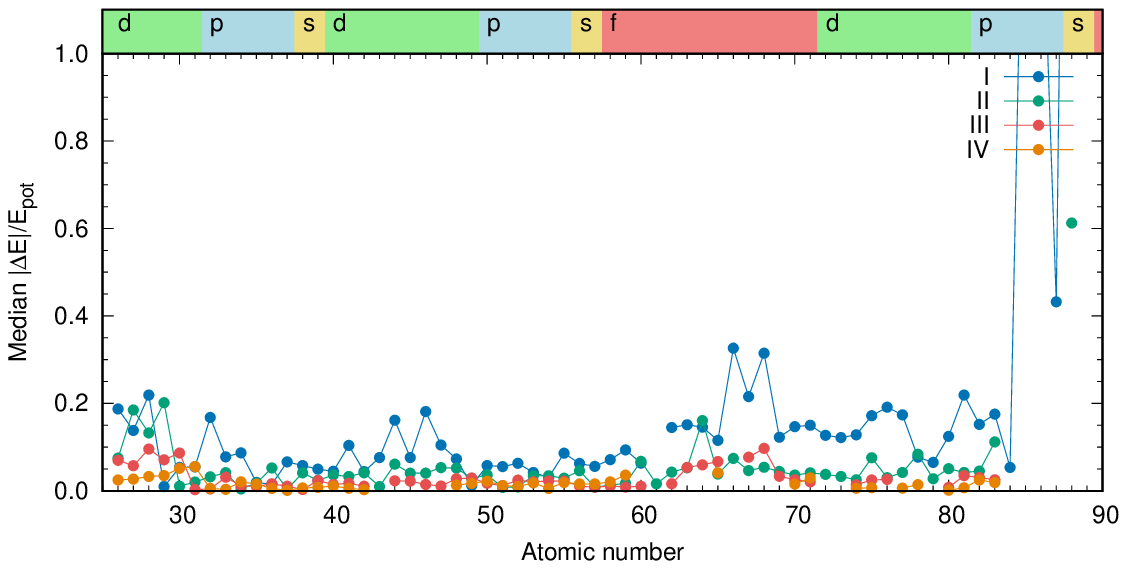}\\
    \caption{
      Comparison of the lowest energy of each configuration
      between our calculations and the NIST data \citep{kramida18}.
      {\it Top}: The number of configurations used for comparison.
      {\it Bottom}: Median of $|\Delta E|/E_{\rm pot}$, where $\Delta E$ is the difference in the lowest energy for a configuration between our calculations and the NIST data and $E_{\rm pot}$ is the ionization potential.
  \label{fig:deltae}
}
\end{center}
\end{figure}
%%%%%%%%%%%%%%%%%%%%%%%%%%%%%%%%%%%%%%%%%%%%%%%%%%%%

Since the calculations involve several assumptions as described above,
we test the validity of our results by comparing the
calculated ionization potential and derived energy levels with
those in the NIST ASD.
Although some uncertainties exist in the NIST data,
this is the best possible way to evaluate our results.
We do not include actinide elements because the energy levels are
poorly known for most of them.

Figure \ref{fig:epot} shows comparison of ionization potentials.
Our atomic calculations give reasonable agreement,
capturing the trend as a function of atomic number.
In general, the agreement in higher ionization states is better,
and the results of the neutral atoms shows the largest deviation
in particular at high atomic numbers ($Z \gsim 60$).
The averaged fractional accuracies as compared with the NIST data
are 14 \%, 7 \%, 4 \%, and 4\% for neutral atoms, singly, doubly,
and triply ionized ions, respectively.

Overall agreement in the energy levels is similar to
our previous results for selected elements \citep{tanaka18}.
Some examples of energy levels are shown in Appendix A.
Figure \ref{fig:deltae} shows typical accuracy of the
lowest energy level for each configuration.
For each configuration, we evaluate the difference in
the lowest energy between our calculations and NIST data ($|\Delta E|$).
The lower panel shows the median of $|\Delta E|$ normalized
by the ionization potential ($E_{\rm pot}$).
The number of configuration
used for the comparison is shown in the upper panel.
As shown in the figure, a typical $|\Delta E|/E_{\rm pot}$
  is $< 20$ \% for neutral and $< 10$ \% for singly to triply ionized ions
   (except for $Z \ge 85$, see Section \ref{sec:elevel}).

It is important to understand how accuracies in atomic calculations
  influence the bound-bound opacities.
  Since the NIST database only includes critically evaluated
  energy levels, the information of energy levels are not complete,
  and thus, we cannot compare the accuracies of all the energy levels.
  Therefore, it is not possible to directly evaluate the impact
  of the accuracy in atomic calculations to the opacities.
  Only the possible way is comparing the opacities calculated 
  with different atomic codes or different assumptions in the atomic calculations.
  \citet{kasen13} shows that
  the difference in the strategy in the atomic calculations results in
  the difference in the opacity of Nd up to by a factor of 2.
  \citet{tanaka18} and \citet{gaigalas19} calculated 
  the bound-bound opacities of selected $r$-process elements 
  by using the HULLAC code and the GRASP2K code \citep{jonsson13},
  which enables more {\it ab-initio} calculations without free parameters.
  They found that (1) overall wavelength dependence of the bound-bound
  opacities agree very well,
  (2) the maximum deviation is about a factor of 2 in
  ultraviolet wavelengths, and (3) the difference in the Planck
  mean opacities is up to a factor of 1.5.
  A similar agreement, within a factor of 1.5-2,
  is also seen in the opacities of \ion{Nd}{ii}
  calculated by \citet{kasen13} and \citet{fontes17}.

Therefore, we regard that a typical level of systematic uncertainty
in the bound-bound opacity is about a factor of 2.
As shown in the following sections,
the elemental variation of the opacities 
is much larger than this uncertainty.
This means that lack of data for some elements have
a bigger impact to the opacity than the accuracy
and systematics in the atomic calculations.

%      It is, however, noted that extensive comparison
%    has been performed only for selected ions
%    because systematic, {\it ab-initio} calculations
%    for many ions are still not feasible.

%%%%%%%%%%%%%%%%%%%%%%%%%%%%%%%%%%%%%%%%%%%%%%%%%%%%
% Figure
%%%%%%%%%%%%%%%%%%%%%%%%%%%%%%%%%%%%%%%%%%%%%%%%%%%%
\begin{figure*}
  \begin{center}
    \includegraphics[scale=1.0]{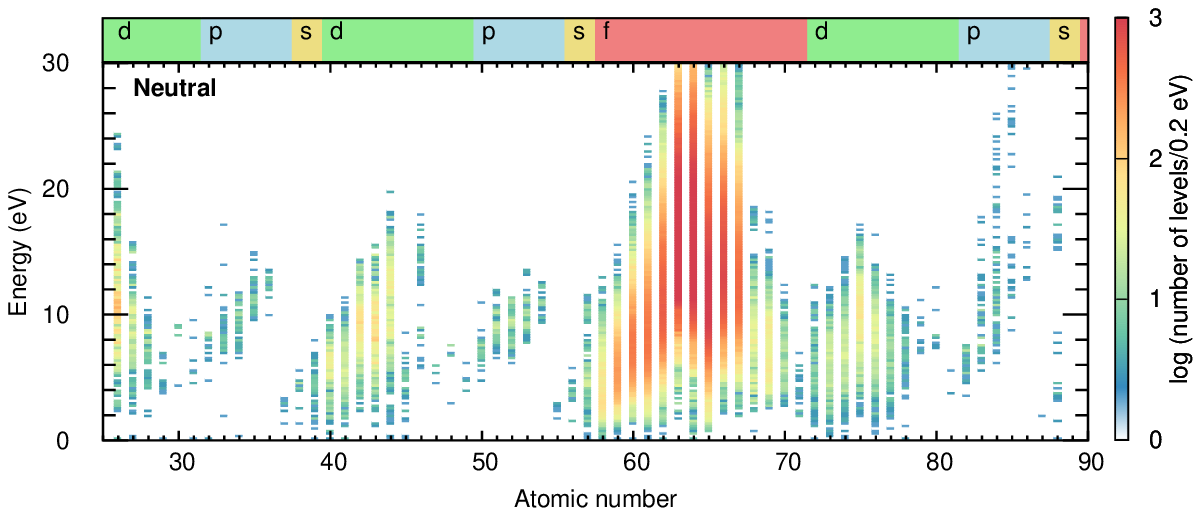}\\
    \includegraphics[scale=1.0]{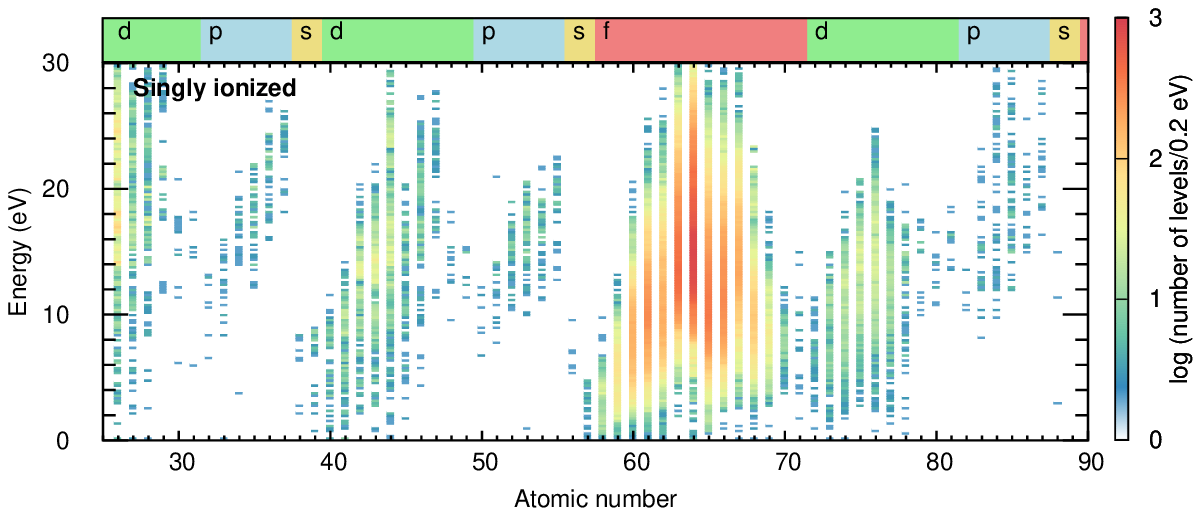}\\
    \includegraphics[scale=1.0]{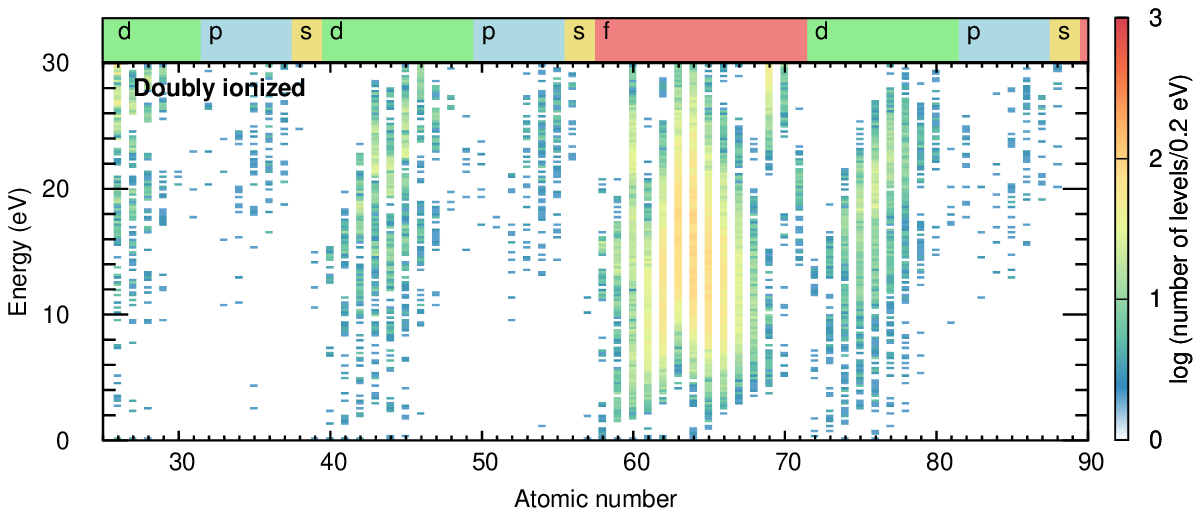}\\
    \includegraphics[scale=1.0]{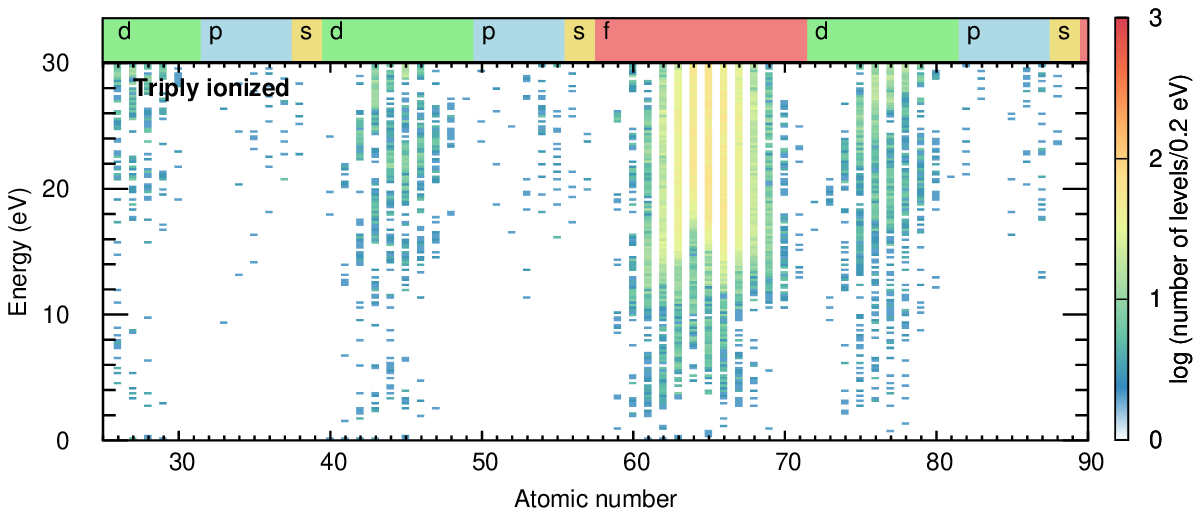}         
\caption{
  \label{fig:elevel}
  Distribution of energy levels of all the elements
  (neutral atom to triply ionized ion from top to bottom panels).
  The color scale represents the number of energy levels
  in 0.2 eV energy bin.}
\end{center}
\end{figure*}
%%%%%%%%%%%%%%%%%%%%%%%%%%%%%%%%%%%%%%%%%%%%%%%%%%%%

%%%%%%%%%%%%%%%%%%%%%%%%%%%%%%%%%%%%%%%%%%%%%%%%%%%% 
% Figure: Planck mean opacity for all the elements
%%%%%%%%%%%%%%%%%%%%%%%%%%%%%%%%%%%%%%%%%%%%%%%%%%%% 
\begin{figure*}
  \begin{center}
    \includegraphics[scale=1.2]{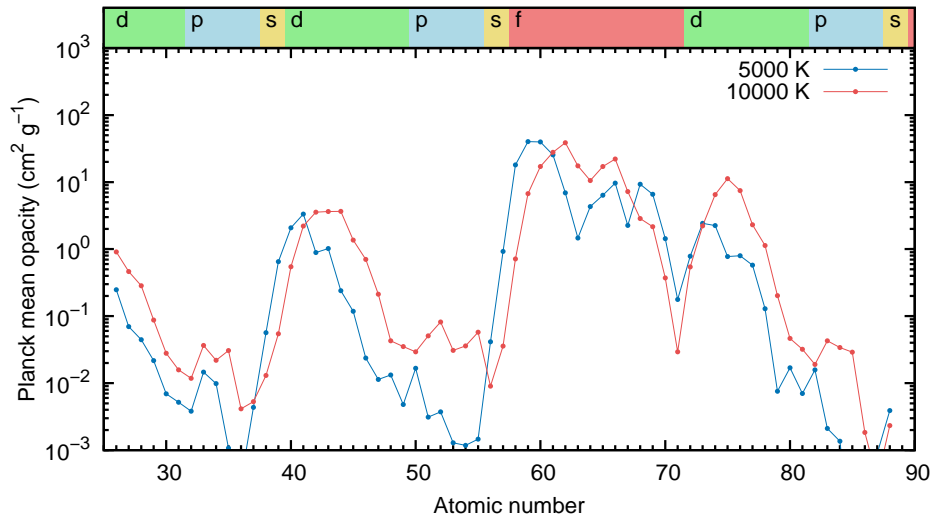} 
\caption{
  \label{fig:opacity}
  Planck mean opacities for all the elements.
  The opacities are calculated by assuming 
  $\rho = 1 \times 10^{-13} \ {\rm g \ cm^{-3}}$, 
  and $t=$ 1 day after the merger.
  Blue and red lines present the opacities for $T = 5,000$ and 10,000 K,
  respectively.
}
\end{center}
\end{figure*}
%%%%%%%%%%%%%%%%%%%%%%%%%%%%%%%%%%%%%%%%%%%%%%%%%%%%

\subsection{Energy levels}
\label{sec:elevel}

Figure \ref{fig:elevel} summarizes the calculated energy levels
for all the elements from $Z=26$ to $Z=88$.
The color scale represents the distribution of energy levels, i.e.,
the number of energy levels in every 0.2 eV energy bin.
As expected from complexity measure \citep{kasen13},
$f$-shell elements have a larger number of energy levels
than the other elements and then $d$-shell and $p$-shell elements follow.

Thanks to the systematic calculations for many elements,
we identify the following two effects that mainly
determine the energy level distribution.
%The trend of the energy levels is determined by
%the combination of two effects as follows.
(1) Within a certain electron shell,
the distribution of the energy levels
tend to be shifted toward higher energy as more electrons occupy the shell
(\eg $Z=40$-$48$ for the case of $4d$ shell
and $Z=57$-$71$ for the case of $4f$ shell).
Since orbital radii become smaller with $Z$,
Coulomb and spin-orbit integrals increase for higher $Z$,
or in other words, electron-electron interaction and
spin-orbit interaction energies become higher for higher $Z$.
Therefore, the energy spacing,
\ie the energy difference to the neighboring level
with the same parity and total angular momentum states,
also increases along with $Z$ for a particular shell
(see Figure 20-2 of \citealt{cowan81} for the case of $4f$ shell).
As a result, the distribution of the energy levels becomes
wider for higher $Z$ for a given shell.
(2) At the same time, the number of states is the largest
for the half-closed shell since it gives the highest complexity,
\ie the number of combinations formed from different
quantum numbers is the largest.

The latter effect is strong in lanthanide elements ($Z = 57-71$).
The total number of levels is the largest for elements around Eu or Gd
which have half closed $4f$-shells.
But the distribution of the energy levels is pushed up for these elements,
and thus, the number of low-lying levels is not necessarily
higher than that of other lanthanide elements.
This is the reason why the bound-bound opacities
of these complex elements are not always higher than
those of the other lanthanides (see Section \ref{sec:opacity}).
In addition, due to the former effect,
the energy distributions of the elements with the conjugate configurations
are wider for higher $Z$.
For example, Dy I ($Z=66$, with the ground configuration of
$4f^{10}6s^1$) has a wider energy distribution
than the conjugate Nd I ($Z=60$, $4f^46s^1$).
By this effect, Nd I tends to have higher bound-bound opacities
than Dy I, in particular for a low temperature (see Section \ref{sec:opacity}).

  In Figure \ref{fig:deltae}, the elements with $Z \ge 85$ show large
  deviations from the NIST data.
  For these elements, configuration energies for 6s, 6p and 7s electrons
  are found to be pushed up
  because electrons in these orbitals are too much compressed
  in a small region, and feel a strong electron-electron repulsion.
%  In fact, Figure \ref{fig:elevel} shows that 
%  the energy level distributions for 6p elements with $Z=81-66$
%  tend to be higher than those for 5p elements with $Z=49-54$.  
  The repulsive interaction is effectively reduced
  by including configuration mixing with outer orbitals,
  but the overall energy level distribution tends
  to be extended to higher energy.
  Also, by the presence of the strong mixing configuration,
  the label of the energy level is not clear,
  which makes direct comparison with the NIST data difficult.
  Due to these facts, the opacities of these heavy elements
  can be more uncertain than those of the other elements.
  We discuss the impact to the opacities in the following sections.

%%%%%%%%%%%%%%%%%%%%%%%%%%%%%%%%%%%%%%%%%%%%%%%%%%%% 
% Section: Opacity
%%%%%%%%%%%%%%%%%%%%%%%%%%%%%%%%%%%%%%%%%%%%%%%%%%%% 

\section{Bound-bound opacity}
\label{sec:opacity}

In a typical timescale of kilonova emission ($t \gsim 1$ day),
bound-bound transitions of heavy elements
is the dominant source for the opacities in near ultraviolet, optical,
and infrared wavelengths \citep{kasen13,tanaka13}.
Bound-free and free-free transitions and electron scattering give
  only minor contributions, although they are included in the radiative
transfer simulations shown in Section \ref{sec:kilonova}.
In the ejecta of NS merger or supernovae,
  Sobolev approximation \citep{sobolev60} can be applied 
  as a large velocity gradient exists and 
  the thermal line width ($\sim 1 $ \kms) is negligible
  compared with the expansion velocity \citep{kasen13}.
  The optical depth of one bound-bound transition
  can be expressed by Sobolev optical depth $\tau_l$:
\begin{equation}
  \tau_l = \frac{\pi e^2}{m_e c} f_{l} n_{i,j,k} t \lambda_{l}
\end{equation}
for homologously expanding material ($dr/dv = t$).
Here $n$ is the population of a lower level of the transition
($i$-th element, $j$-th ionization stage, and $k$-th excited level)
and $f_{l}$ and $\lambda_{l}$ are the oscillator strength
and transition wavelength, respectively.
The Sobolev approximation cannot be used
  when the wavelength spacing of the strong lines becomes
  comparable to the thermal width.
  We confirmed that, in the typical condition for kilonova,
  the wavelength spacing is larger than the thermal width by a factor of $>10$,
  and thus, the Sobolev approximation is applicable (see Appendix B).

To evaluate the Sobolev optical depth, we need ionization and excitation
($n_{i,j,k}$).
Our calculations assume local thermodynamic equilibrium (LTE),
and ionization states ($j$) are calculated by solving Saha equation.
Population of excited states follow the Boltzmann distribution,
\ie $n_{i,j,k}/n_{i,j,0} = (g_{k}/g_{0}) \exp(-E_{k}/kT)$,
where $g_0$ is the statistical weight of the ground state
and $g_k$ and $E_k$ are the statistical weight and energy of
an excited level $k$.
By this exponential dependence of the population of excited states,
bound-bound transitions from lower energy levels have
much higher contributions to the total opacities.

To compute the bound-bound opacity for a certain wavelength grid
$\Delta \lambda$,
we adopt expansion opacity formalism,
which is commonly used for supernovae and NS mergers
\citep{karp77,eastman93,kasen06}.
The expansion opacity for the homologously expanding material
is written as follows:
\begin{equation}
  %  \kappa_{\rm exp} (\lambda) = \frac{\alpha_{\rm exp}^{\rm bb} (\lambda)}{\rho}
  \kappa_{\rm exp} (\lambda) 
= \frac{1}{ct \rho} 
\sum_l \frac{\lambda_l}{\Delta \lambda} (1 - e^{- \tau_l}),
\label{eq:kappa}
\end{equation}
where summation is taken over all the transitions within
the wavelength bin $\Delta \lambda$ in radiative transfer simulations.

Since the summation is calculated for all the calculated transitions,
the expansion opacities can depend on the number of calculated transitions.
The number of transitions is limited by the number of configurations
included in the atomic structure calculations.
In other words, too few configurations in atomic structure calculations
can result in the underestimation
of the bound-bound opacities.
To study the convergence in terms of the number of included configurations,
we calculate the opacities by adding configurations one by one.
We confirm that our choice of configurations give a convergence
in the opacities within 10\%.
This is smaller than the expected systematic uncertainty
of the opacity caused by assumptions in the atomic calculations
(by a factor of 1.5-2, see Section \ref{sec:atomic}).
The details of this convergence studies are given in Appendix A.

In this paper, whenever not explicitly mentioned,
the expansion opacities are evaluated at $t = 1$ day after the merger
by assuming density of $\rho = 1 \times 10^{-13} \ {\rm g \ cm^{-3}}$,
which is typical for the ejecta mass of $\Mej \sim 10^{-2} \Msun$
and the ejecta velocity of $v \sim 0.1c$.
In this section, we show the opacity for each element
\ie the opacity is computed by assuming gas purely consisting of one element
to study the elemental variation of the opacity.
In Section \ref{sec:kilonova},
we show the opacities for the mixture of $r$-process elements.

%To evaluate the bound-bound opacities in rapidly expanding medium,
%such as ejecta of supernova or neutron star merger,
%expansion opacities are commonly used \citep{karp77,eastman93,kasen06}.
%In the homologous expansion, the expansion opacity is expressed by
%\begin{equation}
  %  \kappa_{\rm exp} (\lambda) = \frac{\alpha_{\rm exp}^{\rm bb} (\lambda)}{\rho}
%  \kappa_{\rm exp} (\lambda) 
%= \frac{1}{ct \rho} 
%\sum_l \frac{\lambda_l}{\Delta \lambda} (1 - e^{- \tau_l}),
%\end{equation}
%where summation is taken over all the transitions within
%the wavelength bin $\Delta \lambda$ in radiative transfer simulations.
%Here $\tau_l$ is the Sobolev optical depth for each
%bound-bound transition;
%\begin{equation}
%  \tau_l = \frac{\pi e^2}{m_e c} f_{l} n t \lambda_{l},
%\end{equation}
%where $n$ is the number density in a lower level of the transition
%and $f_{l}$ and $\lambda_{l}$ are the oscillator strength and transition wavelength, respectively.

Figure \ref{fig:opacity} shows the overview of
the opacity as a function of atomic number:
the Plank mean opacities are shown for $T=5,000$ and 10,000 K
for all the calculated elements.
The elemental variation is significant, ranging from $\kappa \sim 10^{-3} \ {\rm cm^2 \ g^{-1}}$ to $\kappa \sim 50 \ {\rm cm^2 \ g^{-1}}$.
  A notable feature is that the variation is quite large even for the elements with the same outermost electron shell.
  In the following sections, properties of the opacities
  and physics behind the behaviors are discussed for the elements with each outermost shell.

%%%%%%%%%%%%%%%%%%%%%%%%%%%%%%%%%%%%%%%%%%%%%%%%%%%% 
% Figure: Opacity (f-shell)
%%%%%%%%%%%%%%%%%%%%%%%%%%%%%%%%%%%%%%%%%%%%%%%%%%%% 
\begin{figure*}
  \begin{center}
    \begin{tabular}{cc}
    \includegraphics[scale=0.9]{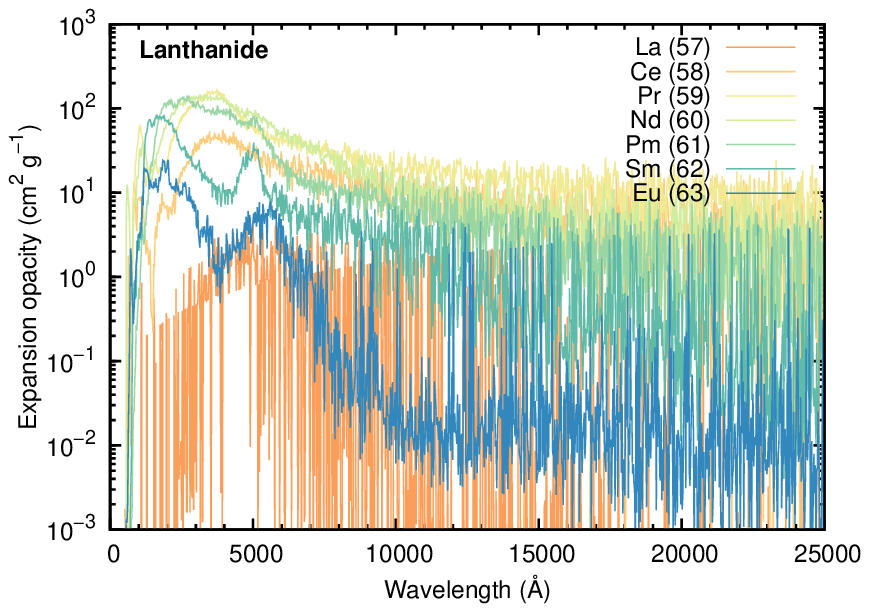} &
    \includegraphics[scale=0.9]{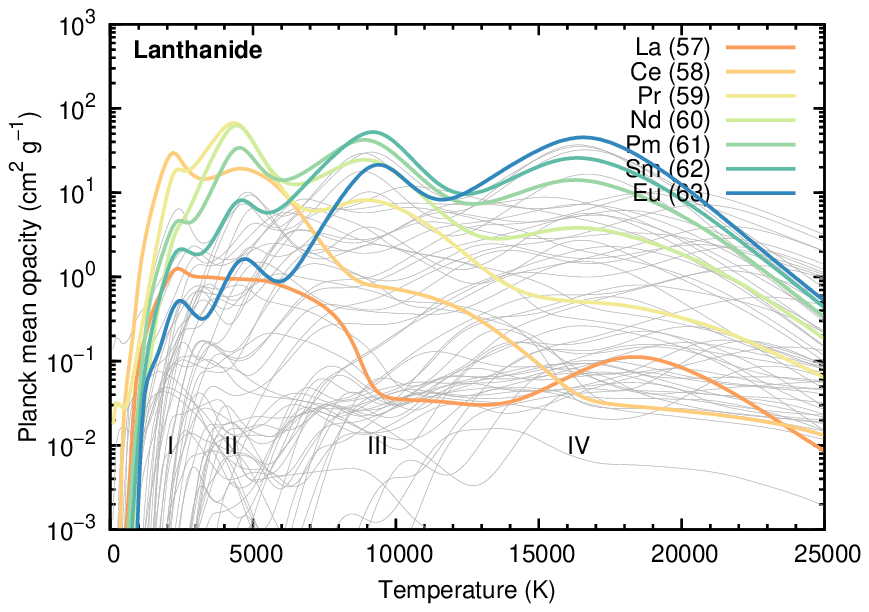}  \\
    \includegraphics[scale=0.9]{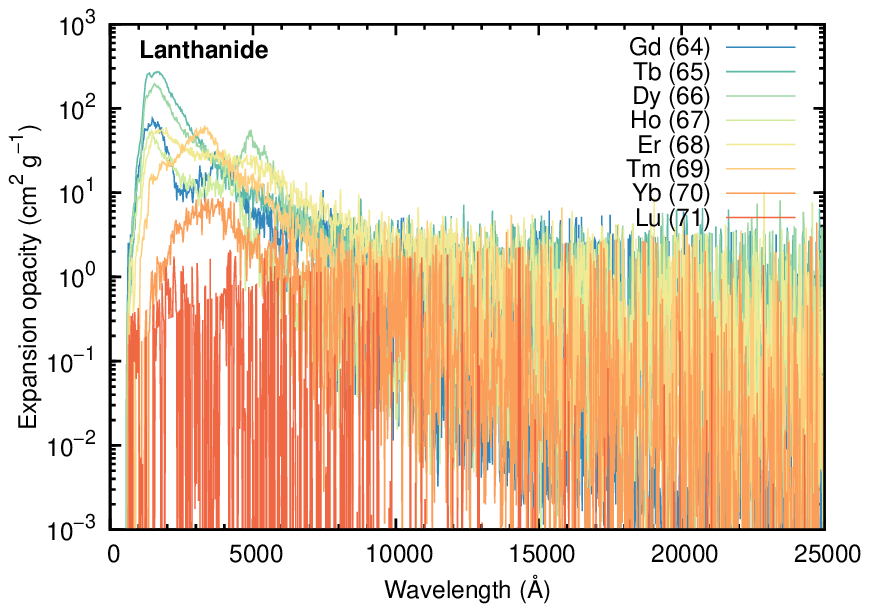} &    
    \includegraphics[scale=0.9]{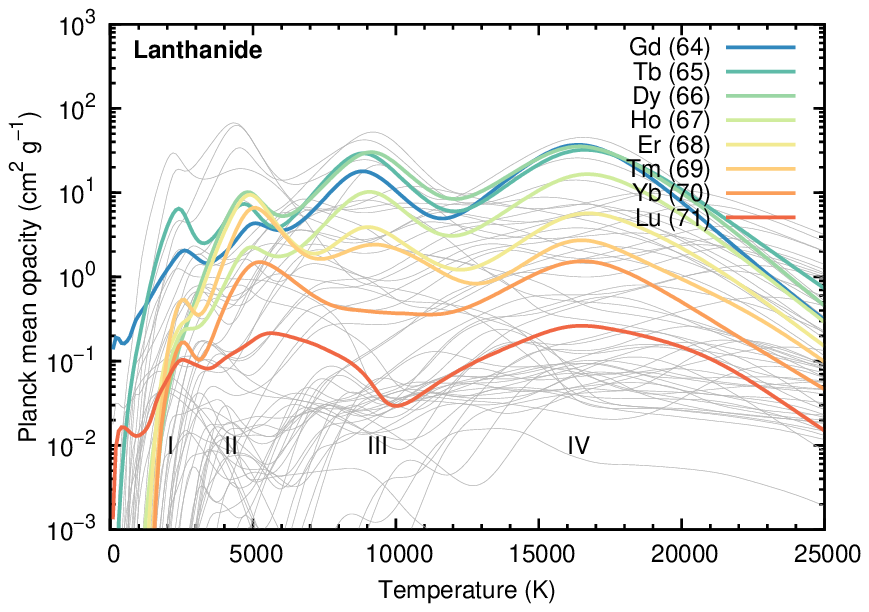}  \\
    \end{tabular}
\caption{
  \label{fig:opacity_f}
  Left: Expansion opacity for $f$-shell (lanthanide) elements
  at $T = 5,000$ K.
  Right: Planck mean opacities as a function of temperature (color).
  Gray lines show the Planck mean opacities
of all the other elements.
The labels (I, II, III, and IV) show typical temperature ranges
  for each ionization state.
}
\end{center}
\end{figure*}
%%%%%%%%%%%%%%%%%%%%%%%%%%%%%%%%%%%%%%%%%%%%%%%%%%%%

%%%%%%%%%%%%%%%%%%%%%%%%%%%%%%%%%%%%%%%%%%%%%%%%%%%% 
% Figure: Opacity vs atomic number
%%%%%%%%%%%%%%%%%%%%%%%%%%%%%%%%%%%%%%%%%%%%%%%%%%%% 
\begin{figure}
  \begin{center}
      \includegraphics[scale=0.9]{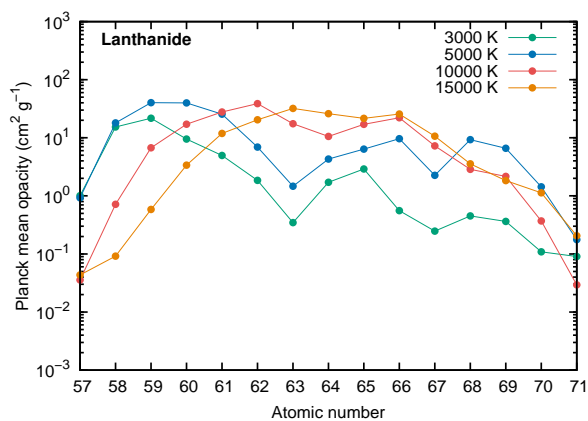} 
\caption{
  \label{fig:opacity_z_la}
  Planck mean opacities of lanthanide elements as a function of atomic number.
  For lower temperature ($T < 5,000$ K), the opacity tends to decrease
  for higher atomic numbers. For higher temperature ($T > 10,000$ K),
  the opacities are highest around half-closed elements.
  }
\end{center}
\end{figure}
%%%%%%%%%%%%%%%%%%%%%%%%%%%%%%%%%%%%%%%%%%%%%%%%%%%%

%%%%%%%%%%%%%%%%%%%%%%%%%%%%%%%%%%%%%%%%%%%%%%%%%%%%
% Subsection: f-shell
%%%%%%%%%%%%%%%%%%%%%%%%%%%%%%%%%%%%%%%%%%%%%%%%%%%%
\subsection{f-shell elements}
\label{sec:f}

Open $f$-shell elements (lanthanides)
have larger opacities than the elements with other outermost electron shells
\citep{kasen13,tanaka13,fontes17,tanaka18,wollaeger18,fontes20}.
Due to the large number of energy levels with small energy spacing,
the opacities remain high in the NIR wavelengths
(left panels of Figure \ref{fig:opacity_f}).
Depending on the elements and temperature,
the Planck mean opacities are $\kappa \sim 0.1 - 50 \ {\rm cm^2 \ g^{-1}}$
(right panels).

For $T=5,000$ K, Planck mean opacities of Pr, Nd, and Pm ($Z=59, 60$, and 61)
are the highest among lanthanide elements
(Figure \ref{fig:opacity_z_la}).
The opacities gradually decrease as more electrons occupy 4$f$-shell.
This is because the number of low-lying energy levels
decreases as $f$-shell has more electrons (\ie $Z$ increases).
Although the total number of energy levels is the largest
for nearly half-closed $f$-shell elements (Eu or Gd),
their opacities are not necessarily highest,
as also found by \citet{kasen17} and \citet{fontes20}.
This is understood by the relatively high energy level distributions
of Eu and Gd (Figure \ref{fig:elevel}).

For $T > 10,000$ K, the Planck mean opacities are the highest
for nearly half-closed elements (Figure \ref{fig:opacity_z_la}).
This is because high excited levels of Eu or Gd
start to contribute to the opacities.
Also, at this temperature, the lanthanides are doubly ionized
and low-$Z$ lanthanide elements
such as Pr and Nd have smaller contributions to the opacities.

Temperature dependence is different for low and high
electron occupations in $f$-shell (Figure \ref{fig:opacity_z_la}).
This dependence is more clearly visible in the right panels
of Figure \ref{fig:opacity_f}.
Low-$Z$ lanthanide elements such as Ce, Pr, Nd ($Z =$ 58, 59, and 60)
show decreasing Planck mean opacities as a function of temperature
because they have smaller number of electrons in 4$f$-shell.
On the other hand, elements with more $f$-shell electrons 
such as Sm, Eu, Gd, Tb, Dy, Ho, Er, Tm, and Yb ($Z = 62-70$)
show increasing opacities with temperature
since they become closer to half-closed shell as temperature increases.

As shown in the right panels of Figure \ref{fig:opacity_f},
our opacity data for $f$-shell elements are applicable only at
$T \lsim 20,000$ K
since our atomic calculations include only up to triply ionized ions.
This temperature corresponds to about $0.5-1$ day after the merger
although this epoch depends on the ejecta parameters such as mass,
velocity, and opacity.
We need atomic calculations for highly ionized ions to correctly understand
the emission at earlier epochs.

%%%%%%%%%%%%%%%%%%%%%%%%%%%%%%%%%%%%%%%%%%%%%%%%%%%% 
% Figure: opacity d-shell
%%%%%%%%%%%%%%%%%%%%%%%%%%%%%%%%%%%%%%%%%%%%%%%%%%%% 
\begin{figure*}
  \begin{center}
    \begin{tabular}{cc}
      \includegraphics[scale=0.9]{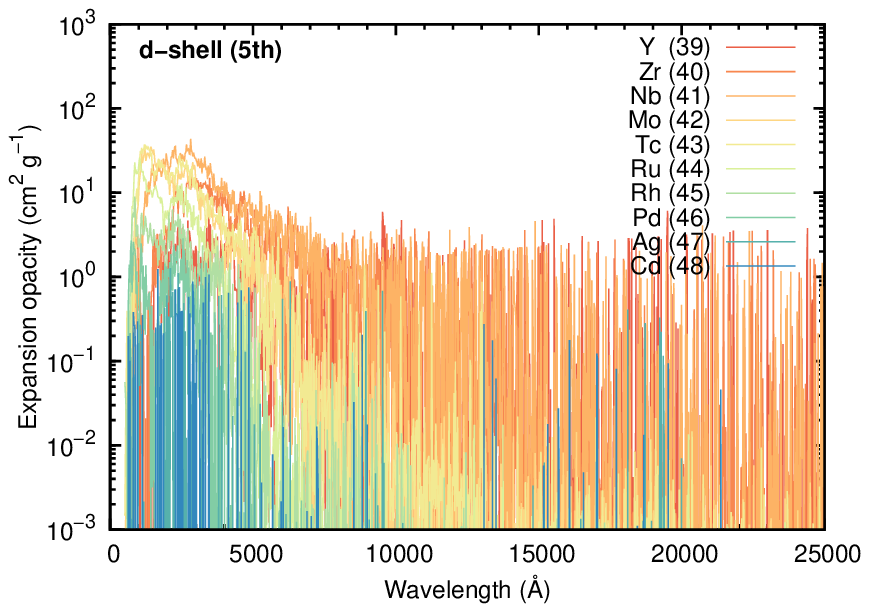} &
      \includegraphics[scale=0.9]{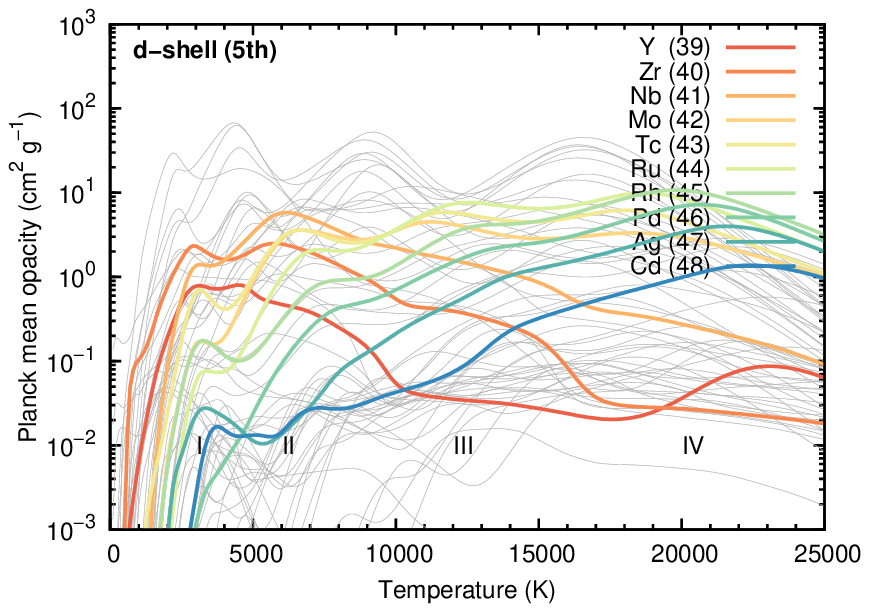} \\
      \includegraphics[scale=0.9]{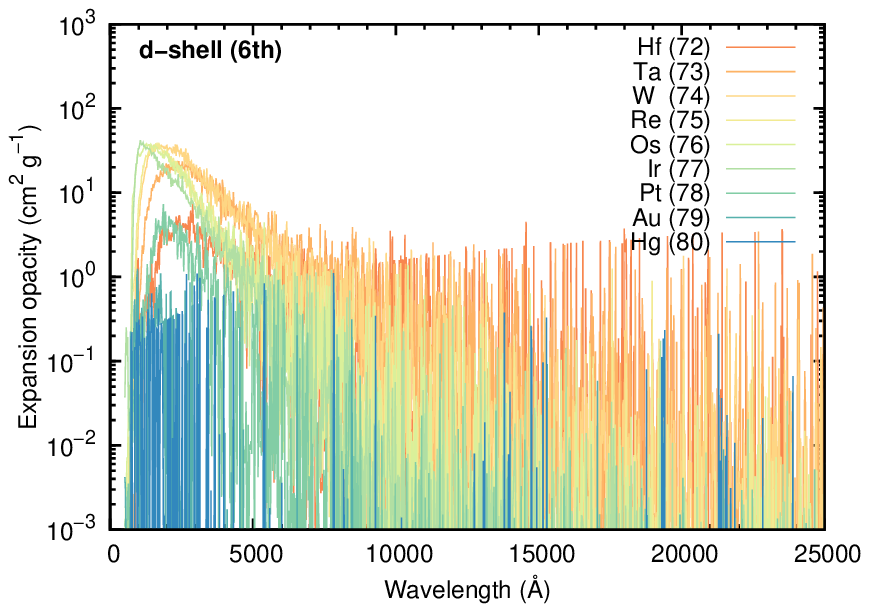} &
      \includegraphics[scale=0.9]{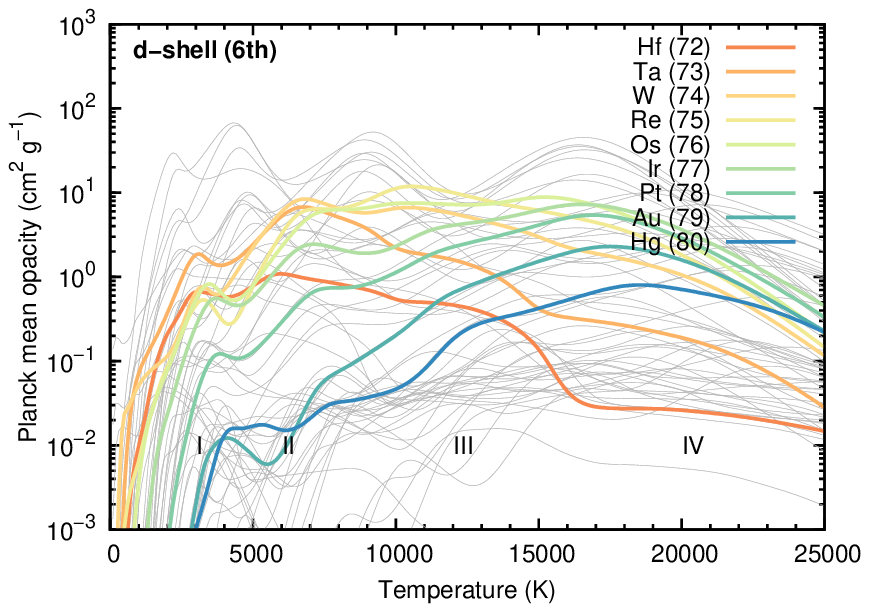}       
    \end{tabular}  
\caption{
  \label{fig:opacity_d}
  Same as Figure \ref{fig:opacity_f} but for $d$-shell elements.
  }
\end{center}
\end{figure*}
%%%%%%%%%%%%%%%%%%%%%%%%%%%%%%%%%%%%%%%%%%%%%%%%%%%%

%%%%%%%%%%%%%%%%%%%%%%%%%%%%%%%%%%%%%%%%%%%%%%%%%%%% 
% Figure: Opacity vs Occupation (group)
%%%%%%%%%%%%%%%%%%%%%%%%%%%%%%%%%%%%%%%%%%%%%%%%%%%% 
\begin{figure}
  \begin{center}
    \includegraphics[scale=0.9]{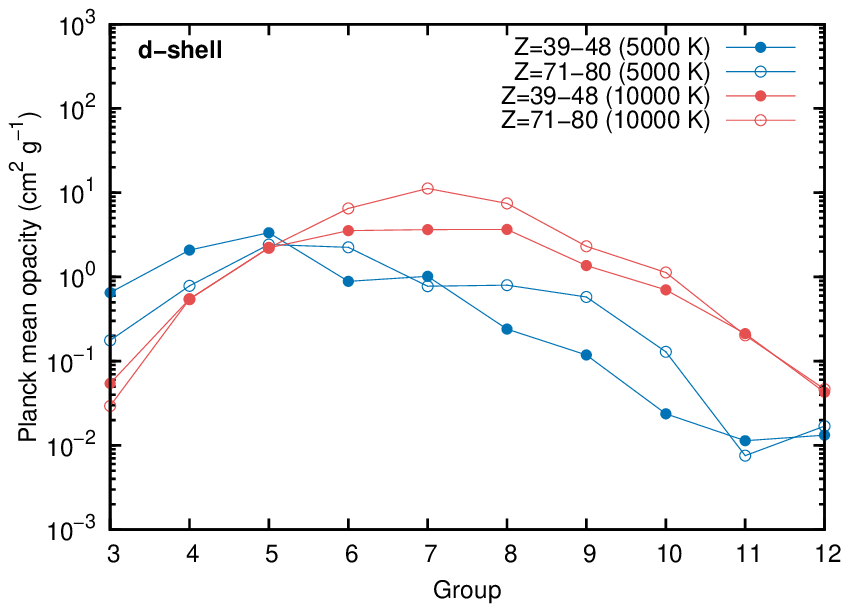}
\caption{
  \label{fig:opacity_z_d}
Planck mean opacities of $d$-shell elements ($Z=39-48$ in the 5th period
  and $Z = 71-80$ in the 6th period) as a function group in the periodic table
  (the group number is approximately the number of electrons
  in the $d$-shell for the case of neutral atoms).}
\end{center}
\end{figure}
%%%%%%%%%%%%%%%%%%%%%%%%%%%%%%%%%%%%%%%%%%%%%%%%%%%%

%%%%%%%%%%%%%%%%%%%%%%%%%%%%%%%%%%%%%%%%%%%%%%%%%%%%
% Subsection: d-shell
%%%%%%%%%%%%%%%%%%%%%%%%%%%%%%%%%%%%%%%%%%%%%%%%%%%%
\subsection{d-shell elements}
\label{sec:d}

Open $d$-shell elements have the second largest contributions
to the opacities after open $f$-shell elements.
Compared with the $f$-shell elements,
the opacities of the $d$-shell elements have a
stronger wavelength dependence, i.e., the opacities
are more concentrated to the shorter wavelengths
around $1,000-3,000$ \AA\ (left panels of Figure \ref{fig:opacity_d}).
The Planck mean opacities are within the range of
 $\kappa \sim 0.01 - 10 \ {\rm cm^2 \ g^{-1}}$ (right panels).

For relatively low temperature ($T < 5,000$ K),
the elements with a smaller number of $d$-shell electrons tend to have
larger opacities (Figure \ref{fig:opacity_z_d}).
This is due to the lower energy level distributions
and larger number of active strong transitions
for the elements with the smaller number of $d$-shell electrons
(Figure \ref{fig:elevel}).
For a higher temperature, the contributions to the opacities
from the elements with 1 or 2 electrons
in neutral atoms (Zr and Nb for 4$d$, Hf and Ta for 5$d$) becomes smaller
 (right panels in Figure \ref{fig:opacity_d})
since these elements do not have $d$-shell electrons when doubly ionized.
This is the reason why the Planck mean opacities have a peak
around groups 7 and 8 at $T = 10,000$ K.

As in the case of $f$-shell elements,
opacities are underestimated at a high temperature ($T \gsim 20,000$ K)
due to the lack of atomic data of higher ionization states.
The applicable temperature range for $d$-shell elements
is wider than that of $f$-shell elements
because of the higher ionization potential of the $d$-shell elements.

%%%%%%%%%%%%%%%%%%%%%%%%%%%%%%%%%%%%%%%%%%%%%%%%%%%% 
% Figure: opacity p-shell
%%%%%%%%%%%%%%%%%%%%%%%%%%%%%%%%%%%%%%%%%%%%%%%%%%%% 
\begin{figure*}
  \begin{center}
    \begin{tabular}{cc}
    \includegraphics[scale=0.9]{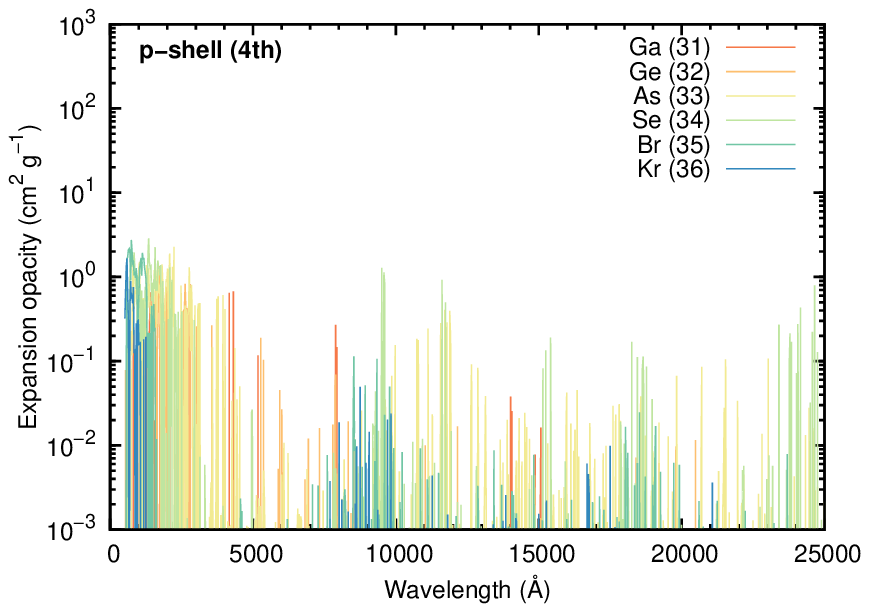} &
    \includegraphics[scale=0.9]{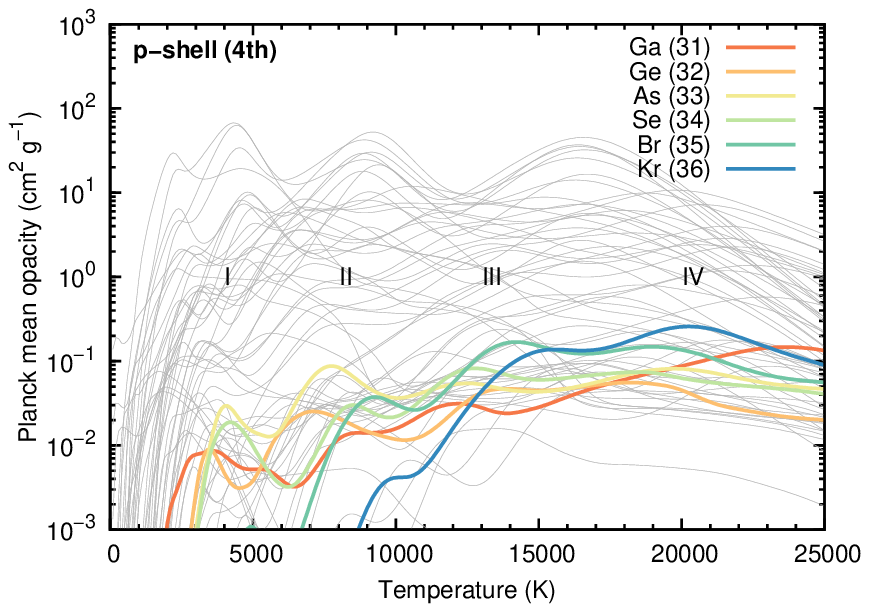} \\
    \includegraphics[scale=0.9]{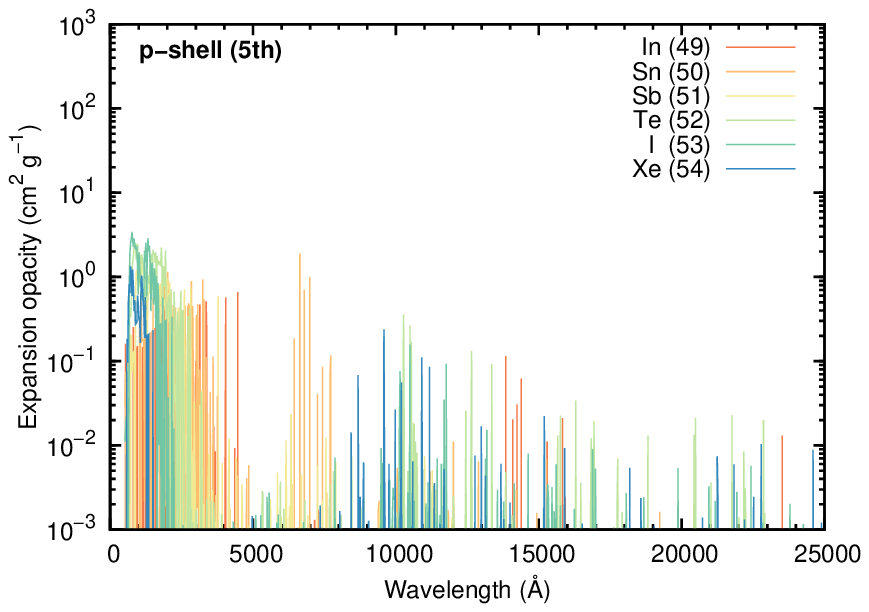} &
    \includegraphics[scale=0.9]{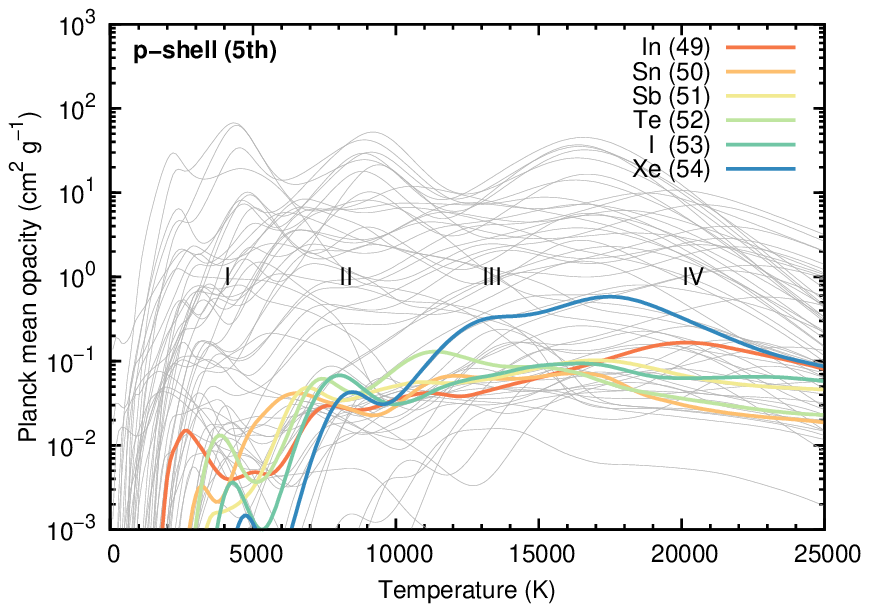} \\
    \includegraphics[scale=0.9]{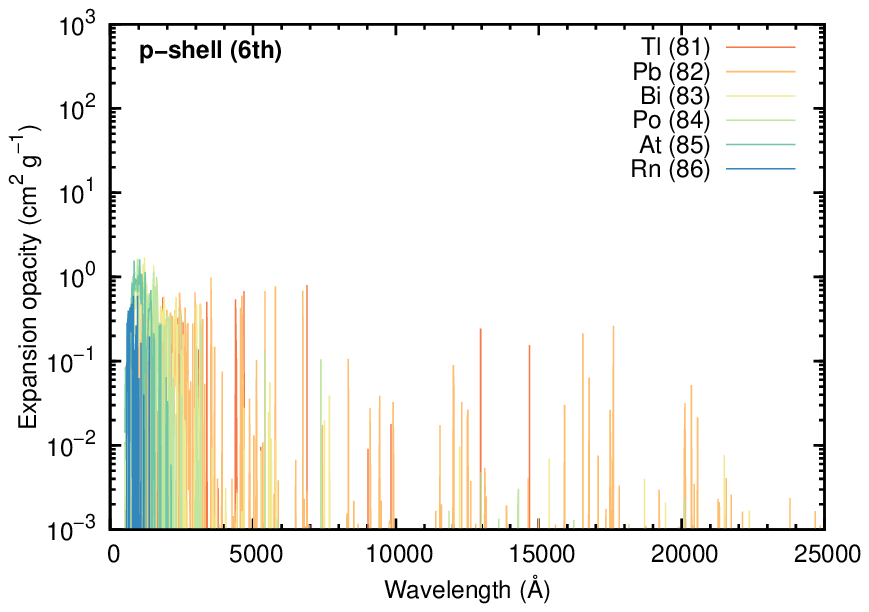} &
    \includegraphics[scale=0.9]{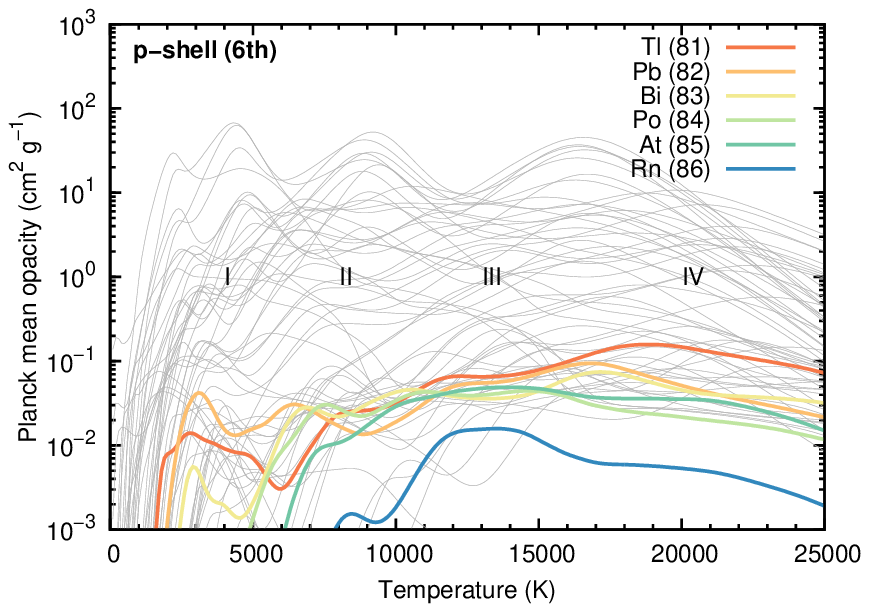}        
    \end{tabular}
\caption{
  \label{fig:opacity_p}
 Same as Figure \ref{fig:opacity_f} but for $p$-shell elements.
}
\end{center}
\end{figure*}
%%%%%%%%%%%%%%%%%%%%%%%%%%%%%%%%%%%%%%%%%%%%%%%%%%%%

%%%%%%%%%%%%%%%%%%%%%%%%%%%%%%%%%%%%%%%%%%%%%%%%%%%%
% Subsection: p-shell
%%%%%%%%%%%%%%%%%%%%%%%%%%%%%%%%%%%%%%%%%%%%%%%%%%%%
\subsection{p-shell elements}
\label{sec:p}

Open $p$-shell elements have smaller contributions to the
opacities compared with open $d$-shell and $f$-shell elements
\citep{kasen13,tanaka18,wollaeger18}.
The opacities are highest at ultraviolet wavelengths
(left panels of Figure \ref{fig:opacity_p}).
For the optical and near-infrared wavelengths,
the Planck mean opacities increase as a function of temperature
but they are at most $\kappa \sim 1  \ {\rm cm^2 \ g^{-1}}$
for $T < 20,000$ K (right panels).

As in the cases for open $d$-shell elements,
the opacities of $p$-shell elements are smaller for more $p$-shell electrons
(Figure \ref{fig:opacity})
since the distribution of energy levels is shifted toward higher energy.
This trend is more significant because the average energy levels
of $p$-shell elements are higher than those of $d$-shell elements
(Figure \ref{fig:elevel}).

  As discussed in Section \ref{sec:atomic},
  the elements with $Z \ge 85$ show large deviation in the energy level
  as compared with the NIST data.
  The energy levels of these elements tend to be pushed up.
  As a result, the opacities of these elements are 
  significantly underestimated.
  At $T < 5000$ K, the Planck mean opacity of At ($Z=85$)
  becomes lower than that of I ($Z=53$) by a factor of 100.
  The effect is even bigger for Rn ($Z=86$):
  the opacity of Rn is lower than that of Xe ($Z=54$) by a factor of 100
  for a wide temperature range.
  Therefore, we regard that the opacities of these elements are
  not reliable.
  Note that the contribution of these elements
  to the total opacity in the NS merger ejecta
  is quite small (see Section \ref{sec:kilonova}).

%%%%%%%%%%%%%%%%%%%%%%%%%%%%%%%%%%%%%%%%%%%%%%%%%%%% 
% Figure: Opacity s-shell
%%%%%%%%%%%%%%%%%%%%%%%%%%%%%%%%%%%%%%%%%%%%%%%%%%%% 
\begin{figure*}
  \begin{center}
    \begin{tabular}{cc}
    \includegraphics[scale=0.9]{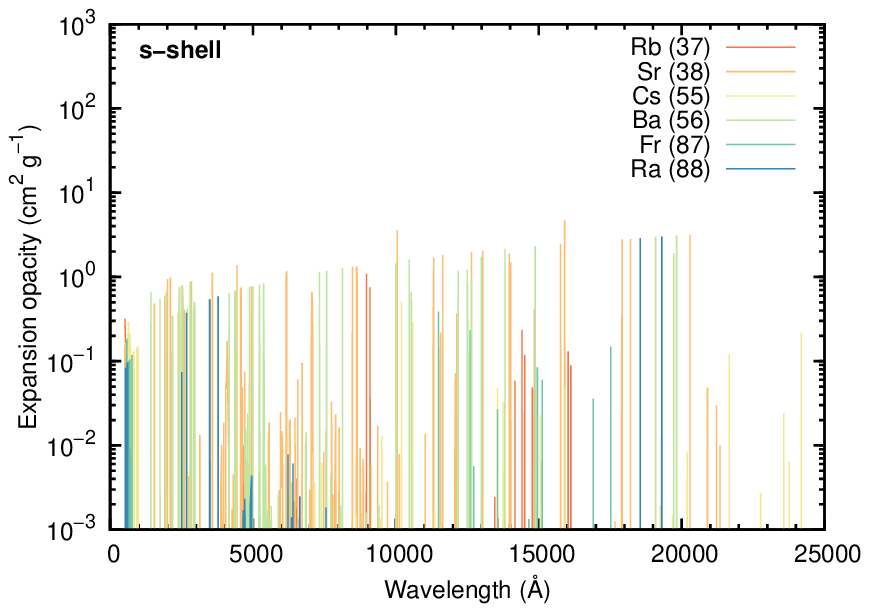} &
    \includegraphics[scale=0.9]{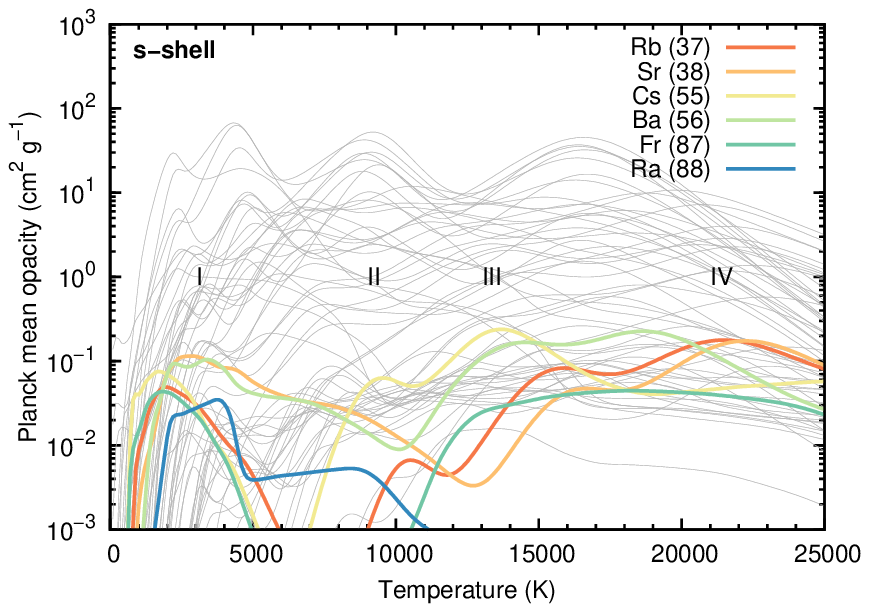} 
    \end{tabular}
\caption{
  \label{fig:opacity_s}
  Same as Figure \ref{fig:opacity_f} but for $s$-shell elements.
  }
\end{center}
\end{figure*}
%%%%%%%%%%%%%%%%%%%%%%%%%%%%%%%%%%%%%%%%%%%%%%%%%%%%

%%%%%%%%%%%%%%%%%%%%%%%%%%%%%%%%%%%%%%%%%%%%%%%%%%%%
% Subsection: s-shell
%%%%%%%%%%%%%%%%%%%%%%%%%%%%%%%%%%%%%%%%%%%%%%%%%%%%
\subsection{s-shell elements}
\label{sec:s}

The opacities of open $s$-shell elements are
almost negligible to the total opacities.
Since there are fewer number of transitions,
they do not form quasi-continuum opacities
(left panel of Figure \ref{fig:opacity_s}).
For the typical temperature of kilonovae,
the Planck mean opacities are $\kappa \lsim 0.1  \ {\rm cm^2 \ g^{-1}}$
(right panel).
Note that, as for the case of At ($Z=85$) and Rn ($Z=86$),
  our opacities for Fr ($Z=87$) and Ra ($Z=88$) are not reliable.
  Although the opacity of Fr shows similar trend with Rb ($Z=37$),
  the opacity of Ra is significantly lower than those of the other $s$-shell elements.

Overall low opacities of $s$-shell elements do not necessarily mean
that they do not contribute to the outcome of kilonova emission.
  In fact, open $s$-shell elements such as Na, Mg, and Ca
  often show strong absorption lines in stellar spectra.
  and thus, $s$-shell elements may contribute to absorption
  lines in the spectra (see \citealt{watson19} for the identification of Sr
  in the spectra of AT2017gfo).
Unfortunately, since the atomic calculations for $r$-process elements
do not give spectroscopic accuracy, i.e., high enough accuracy
to predict the exact wavelengths of each transition,
the usefulness of our opacity data for open $s$-shell elements is limited.

%%%%%%%%%%%%%%%%%%%%%%%%%%%%%%%%%%%%%%%%%%%%%%%%%%%% 
% Section: Kilonova
%%%%%%%%%%%%%%%%%%%%%%%%%%%%%%%%%%%%%%%%%%%%%%%%%%%% 

%%%%%%%%%%%%%%%%%%%%%%%%%%%%%%%%%%%%%%%%%%%%%%%%%%%% 
% Figure: Opacity Ye
%%%%%%%%%%%%%%%%%%%%%%%%%%%%%%%%%%%%%%%%%%%%%%%%%%%% 
\begin{figure*}
  \begin{center}
    \begin{tabular}{cc}
    \multicolumn{2}{c}{\includegraphics[scale=0.9]{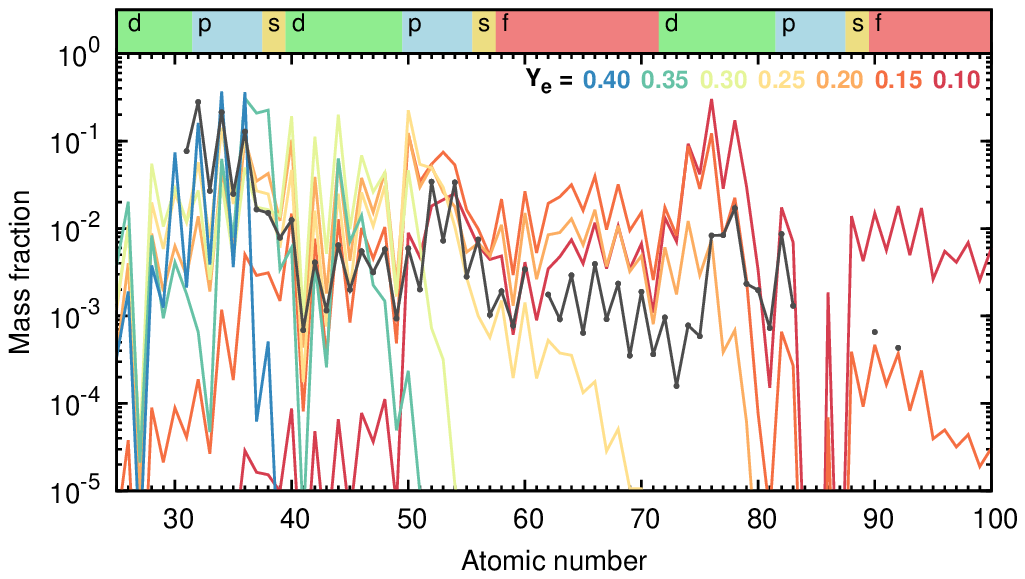}}  \\          
    \includegraphics[scale=0.9]{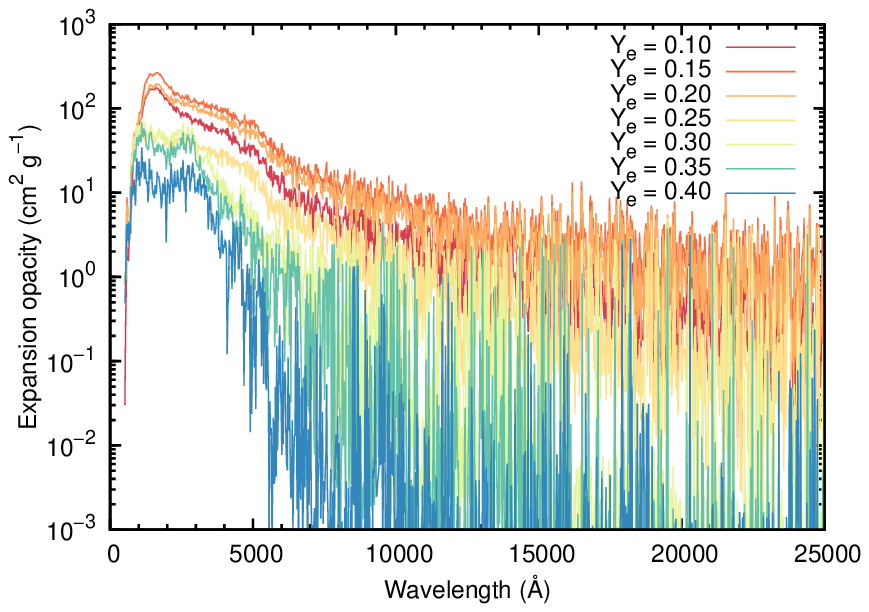} &
    \includegraphics[scale=0.9]{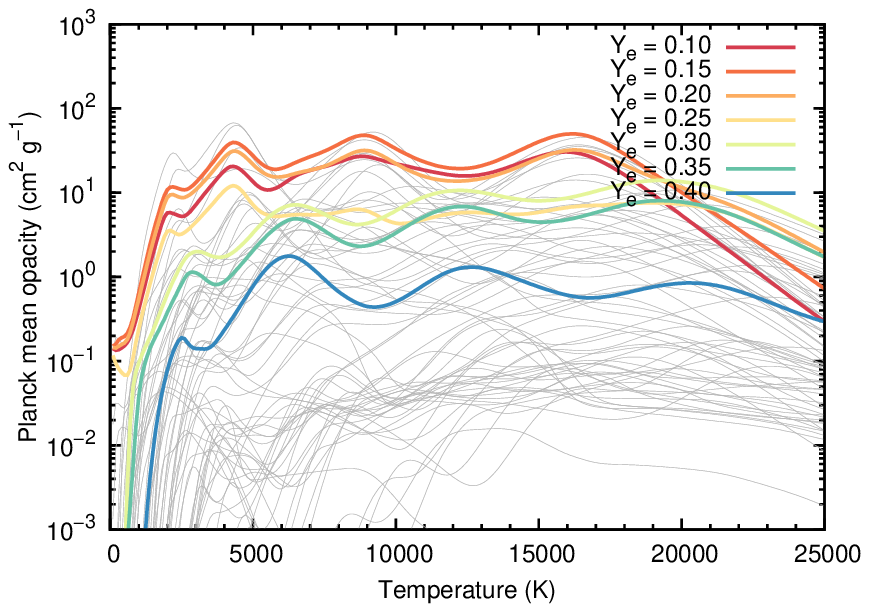} 
    \end{tabular}
\caption{
  \label{fig:opacity_ye}
  Top: Abundance distribution for different $\Ye$ \citep{wanajo14}.
  Bottom left: Expansion opacity as a function of wavelength
  for each $\Ye$.
  Bottom right: Planck mean opacity as a function of temperature
  for each $\Ye$.
  }
\end{center}
\end{figure*}
%%%%%%%%%%%%%%%%%%%%%%%%%%%%%%%%%%%%%%%%%%%%%%%%%%%%

\section{Applications to Kilonovae}
\label{sec:kilonova}

\subsection{Opacities for the mixture of the elements}
\label{sec:mixture}

Ejecta from NS mergers consist of mixture of $r$-process elements.
Abundance distribution is mainly determined by electron fraction $\Ye$.
The first dynamical ejecta have a wide $\Ye$ distribution
down to $\Ye \sim 0.1$ \citep{wanajo14,sekiguchi15,sekiguchi16,goriely15,radice16,foucart16}
while subsequent post-merger ejecta can have higher $\Ye$ 
due to the neutrino absorption if a massive neutron star remains
for longer than a certain period \citep{fernandez13,fujibayashi18,radice18c,fernandez19}.
Lanthanide elements are efficiently produced with $\Ye \lsim 0.25$
\citep[\eg][]{lippuner15,kasen15}.
Therefore, if the ejecta consists of material with $\Ye > 0.25$,
a short-lived, bright and blue emission is expected
due to the absence of high opacity lanthanide
\citep{metzger14,kasen15,tanaka18,wollaeger18,miller19}.

Thanks to the systematic atomic data,
we are now able to connect the abundance pattern or electron fraction $\Ye$
with the atomic opacities in a more reliable manner.
For the mixture of the elements, we construct a line list
from Kurucz's line list \citep{kurucz95} for $Z = 1 - 28$, 
the VALD database \citep{piskunov95,ryabchikova97,kupka99,kupka00}
for $Z = 29$ and 30,
and the results of our new atomic calculations for $Z = 31 - 88$.

Figure \ref{fig:opacity_ye} shows expansion opacities for different $\Ye$.
At $\Ye \le 0.20$, the opacities are dominated by lanthanide elements.
The Planck opacities do not strongly depend on $\Ye$ and 
stay around $\kappa \sim 20-30 \ {\rm cm^2 \ g^{-1}}$ at $T > 5,000$ K.
The temperature dependence at $T > 5,000$ K is weaker than in
individual elements because of the mixture of the elements
with different peaks positions as a function of temperature.
Note that the opacities for the case of $\Ye = 0.10$
may be underestimated due to the lack of actinide elements in our calculations.

These low Ye cases ($\Ye = 0.10$ and $0.15$) synthesize
  the heavy elements ($Z \ge 85$) that show the large deviation
  of the energy levels compared with the NIST data.
  However, the impact to the opacities is limited.
  We calculate the opacities for the mixture of the elements
  by replacing the atomic data for $6p$ and $7s$ elements with
  those for $5p$ and $6s$ elements, respectively.
  We confirm that the Planck mean opacities 
  are affected at most by $1\%$ due to these changes.
  This is reasonable since the mass fractions of these heavy elements are small
  and the opacities of $s$-shell and $p$-shell elements are
  subdominant as compared with $f$-shell elements.
  
The opacities are smaller for higher $\Ye$ as relative fractions
of lanthanides decrease (Table \ref{tab:kappa}).
For $\Ye = 0.25-0.35$, the Planck mean opacities
are dominated by the $d$-shell element (4th period in the periodic table)
and they are in the range of $\kappa = 1 - 10 \ {\rm cm^2 \ g^{-1}}$ at $T > 5,000$ K.
The opacities slightly increase with temperature
due to the contribution of latter half of $d$-shell elements 
(group 8--11, see Figure \ref{fig:opacity_z_d}).
For $\Ye = 0.4$, the contributions from $d$-shell elements decrease
and the opacities are even lower, \ie
$\kappa = 0.1 - 1 \ {\rm cm^2 \ g^{-1}}$ at $T > 5,000$ K.

At a high temperature ($T \gsim 20,000$ K),
the opacity of the low $\Ye$ case decreases more rapidly
than that of the high $\Ye$ case.
This is due to the limitation of ionization states
in our atomic data (see Sections \ref{sec:f} and \ref{sec:d}), \ie
our opacity data are not applicable for high temperature.
Since the ionization potentials of $d$-shell elements are
generally higher than those of $f$-shell elements,
the applicable temperature range is wider for high $\Ye$ cases,
where $d$-shell elements dominate the opacities.

Note that the opacity of $\kappa = 0.1-0.5 \ {\rm cm^2 \ g^{-1}}$ is often used
for blue kilonovae because it gives a good approximation for Type Ia supernova,
where Fe is the major component in the abundance.
However, the opacities of mixture of $r$-process elements
are almost always higher than $\kappa = 0.1-0.5 \ {\rm cm^2 \ g^{-1}}$
even for high $\Ye$, except for a low temperature ($T < 2,000$ K).
This is because Fe is not necessarily representative of $d$-shell elements
and the contribution of Fe-like elements (Ru and Os) is
low compared with other $d$-shell elements
at $T < 10,000$ K (Figure \ref{fig:opacity_d}).

For the ease of applications in analytical models,
we give average values of the Planck mean opacities in Table \ref{tab:kappa}.
%\ref{tab:kappa}.
However, it is emphasized that the average opacities
are derived only at $T = 5,000-10,000$ K
and there is a strong temperature dependence at $T < 5,000$ K.

%%%%%%%%%%%%%%%%%%%%%%%%%%%%%%%%%%%%%%%%%%%%%%%%%%%% 
% Table: Planck mean opacity %%%%%%%%%%%%%%%%%%%%%%%
%%%%%%%%%%%%%%%%%%%%%%%%%%%%%%%%%%%%%%%%%%%%%%%%%%%% 
\begin{table}
  \centering
  \caption{Planck mean opacity for the mixture of the elements.  $\Ye$ is electron fraction,  $X$(La) is mass fraction of lanthanide elements, $X$(La+Ac) is mass fraction of lanthanide and actinide elements, and $\kappa$ is average Planck mean opacity for $T = 5,000-10,000$ K ($\rho = 1 \times 10^{-13} \ {\rm g \ cm^{-3}}$ 
  and $t=$ 1 day after the merger). The opacity shown with $*$ is underestimated due to the lack of complete atomic data for actinide elements.}
  \label{tab:kappa}
  \begin{tabular}{llll}
    \hline    
 $\Ye$    & $X$(La)  &  $X$(La+Ac)  & $\kappa$  \\
    &              &                  & ${\rm cm^2 \ g^{-1}}$ \\
    \hline        
0.10      &  $7.1 \times 10^{-2}$ & $1.7 \times 10^{-1}$ & 19.5$*$ \\
0.15      &  $2.6 \times 10^{-1}$ & $2.6 \times 10^{-1}$ & 32.2 \\
0.20      &  $1.1 \times 10^{-1}$ & $1.1 \times 10^{-1}$ & 22.3 \\
0.25      &  $5.5 \times 10^{-3}$ & $5.5 \times 10^{-3}$ & 5.60 \\
0.30      &  $3.4 \times 10^{-7}$ & $3.4 \times 10^{-7}$ & 5.36   \\
0.35      &  0.0                 &  0.0                & 3.30 \\
0.40      &  0.0                 &  0.0                & 0.96 \\ \\
0.10-0.20 &  $2.1 \times 10^{-1}$ &$2.3 \times 10^{-1}$ & 30.7 \\
0.20-0.30 &  $4.8 \times 10^{-2}$ &$4.8 \times 10^{-2}$ & 15.4 \\
0.30-0.40 &  0.0                 &  0.0                & 4.68 \\
    \hline        
  \end{tabular}
\end{table}
%%%%%%%%%%%%%%%%%%%%%%%%%%%%%%%%%%%%%%%%%%%%%%%%%%%% 

%%%%%%%%%%%%%%%%%%%%%%%%%%%%%%%%%%%%%%%%%%%%%%%%%%%% 
% Figure: Kappa vs time
%%%%%%%%%%%%%%%%%%%%%%%%%%%%%%%%%%%%%%%%%%%%%%%%%%%% 
\begin{figure}
  \begin{center}
    \includegraphics[scale=1.0]{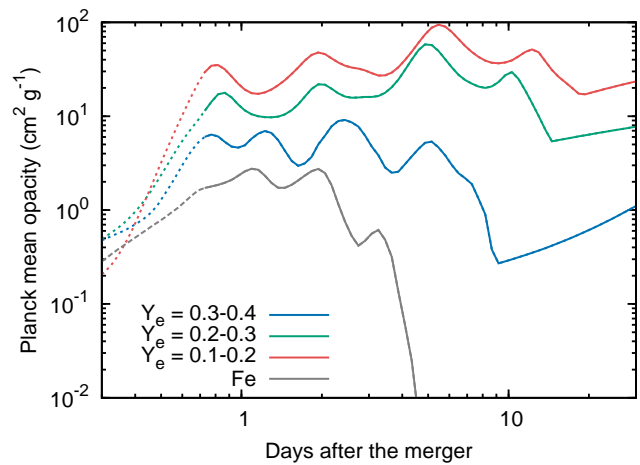}
    \caption{
      Time evolution of the Planck mean opacities
      at $v=0.1c$ in the ejecta
      for the models with high $\Ye$ ($\Ye = 0.30-0.40$, blue),
      intermediate $\Ye$ ($\Ye = 0.20-0.30$, green),
      and low $\Ye$  ($\Ye = 0.10-0.20$, red).
      The gray line shows the opacity of a model with pure Fe ejecta.
  \label{fig:kappa_time}
}
\end{center}
\end{figure}
%%%%%%%%%%%%%%%%%%%%%%%%%%%%%%%%%%%%%%%%%%%%%%%%%%%%

%%%%%%%%%%%%%%%%%%%%%%%%%%%%%%%%%%%%%%%%%%%%%%%%%%%% 
% Figure: Light curves
%%%%%%%%%%%%%%%%%%%%%%%%%%%%%%%%%%%%%%%%%%%%%%%%%%%% 
\begin{figure}
  \begin{center}
    \includegraphics[scale=1.0]{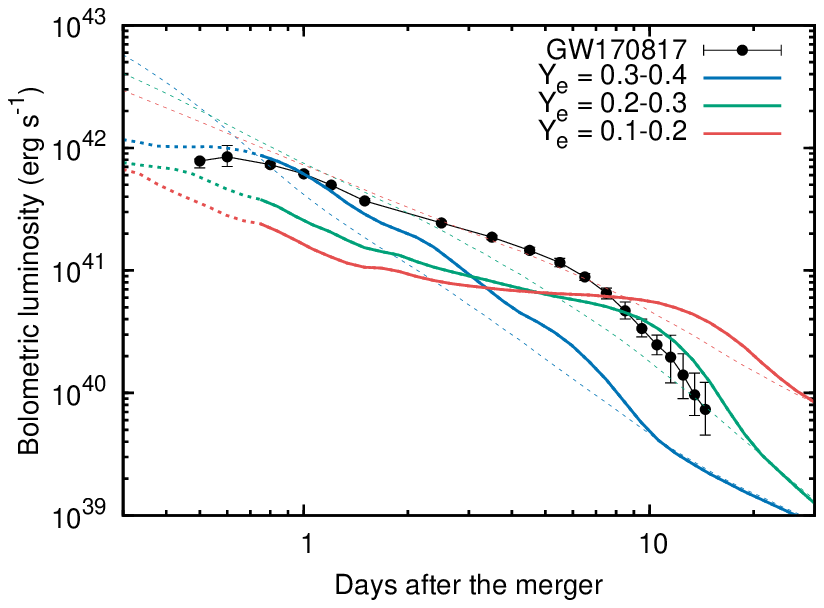}
    \caption{
  Bolometric light curves of the models with high $\Ye$
  ($\Ye = 0.30-0.40$, blue), intermediate $\Ye$ ($\Ye = 0.20-0.30$, green),
  and low $\Ye$  ($\Ye = 0.10-0.20$, red)
  compared with the bolometric light curve of GW170817/AT2017gfo
  constructed by \citet{waxman18}.
  Dotted lines show the epoch in which our calculations are not reliable
  since the ejecta temperature is too high (T $\gsim$ 20,000 K)
    for our opacity data
    (only up to triply ionized ions, see Section \ref{sec:opacity}).
    Thin dashed lines show the luminosity deposited to the ejecta
      (radioactive power multiplied by thermalization efficiency)
      for each model.
    }
  \label{fig:lbol}
\end{center}
\end{figure}
%%%%%%%%%%%%%%%%%%%%%%%%%%%%%%%%%%%%%%%%%%%%%%%%%%%%

%%%%%%%%%%%%%%%%%%%%%%%%%%%%%%%%%%%%%%%%%%%%%%%%%%%% 
% Figure: Light curves
%%%%%%%%%%%%%%%%%%%%%%%%%%%%%%%%%%%%%%%%%%%%%%%%%%%% 
\begin{figure}
  \begin{center}
    \includegraphics[scale=1.0]{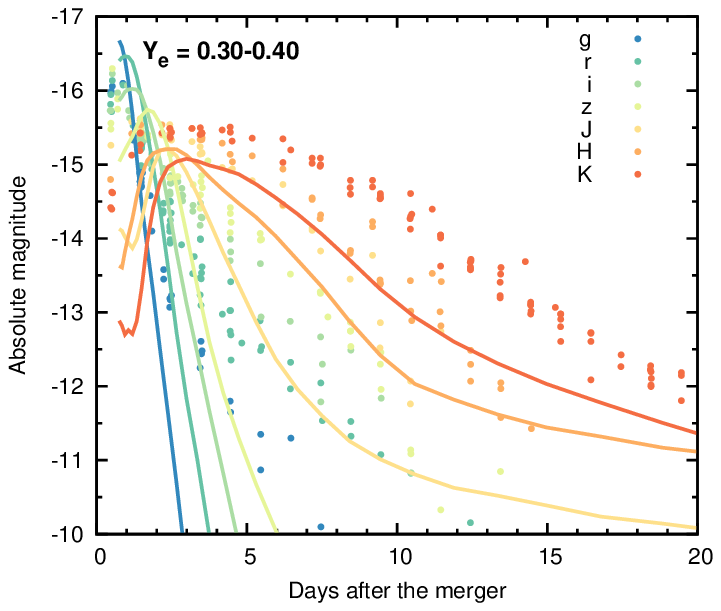}
    \includegraphics[scale=1.0]{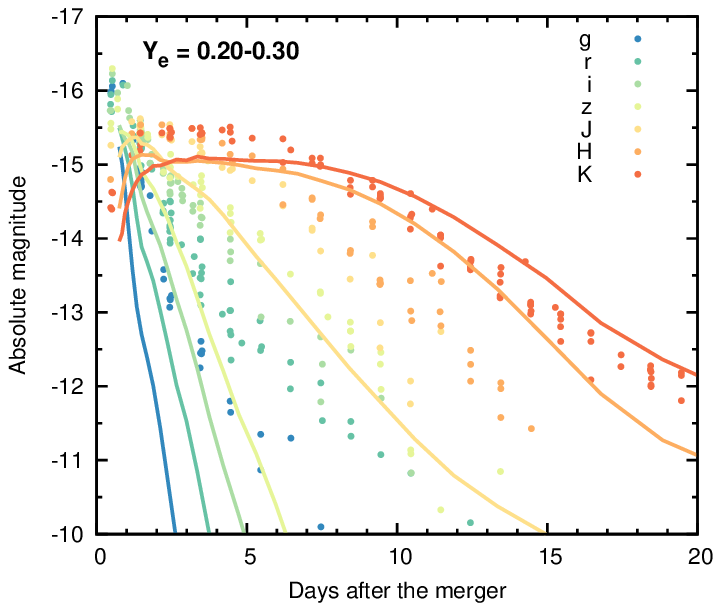}
    \includegraphics[scale=1.0]{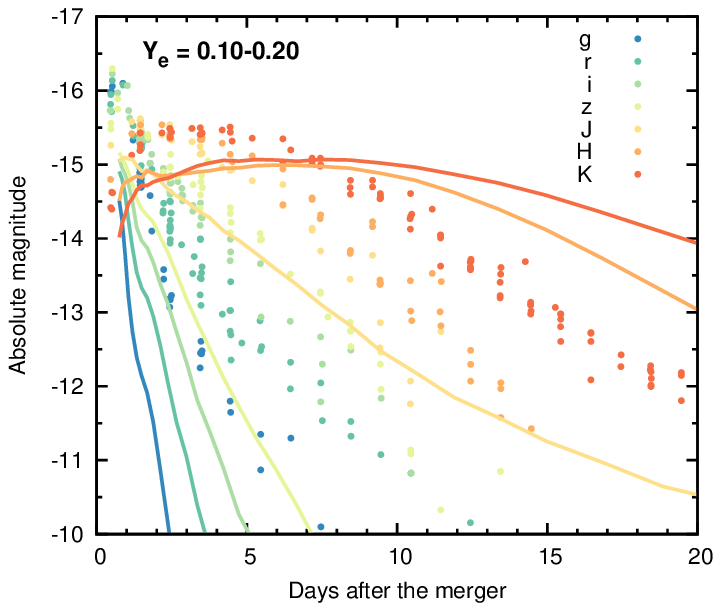}
\caption{
  Multi-color light curves in optical ($griz$) and NIR ($JHK$) filters
  for the models with high $\Ye$ ($\Ye = 0.30-0.40$, top),
  intermediate $\Ye$ ($\Ye = 0.20-0.30$, middle),
  and low $\Ye$  ($\Ye = 0.10-0.20$, bottom)
  compared with the observed light curves of GW170817/AT2017gfo
  (compiled by \citealt{villar17}).
  \label{fig:lc}
}
\end{center}
\end{figure}
%%%%%%%%%%%%%%%%%%%%%%%%%%%%%%%%%%%%%%%%%%%%%%%%%%%%

\subsection{Time evolution of the opacity}
\label{sec:transfer}

The opacities in the NS merger ejecta depend not only
 on elements and temperature but also
 the density of the ejecta (and thus, the position in the ejecta).
 Therefore, the opacities evolves with time by the combination
 of these effects.
In this section, we apply our new atomic data to radiative transfer
simulations of kilonovae
and study the time variation of the opacities in the ejecta.
We use a Monte-Carlo radiative transfer code developed by
\citet{tanaka13,tanaka14} and further updated by \citet{kawaguchi18}
to include special-relativistic effects.
We adopt a simple one-dimensional ejecta model
with a power-law density structure $\rho \propto r^{-3}$
from $v=0.05c$ to $v=0.2c$ \citep{metzger10,metzger17},
which gives an average velocity of $\langle v \rangle = 0.1c$.
The total mass is set to be $\Mej = 0.03 \Msun$.

The radiative transfer code adopts
nuclear heating rates and abundances of $r$-process elements
according to the value of $\Ye$.
Note that, for relatively high $\Ye$,
the nuclear heating rate strongly depends on $\Ye$
since a few isotopes can dominate the heating rate in a certain timescale.
Thus, an assumption of single $\Ye$ ejecta,
which is not the case in realistic conditions, can lead to misleading results.
Therefore, we perform simulations with the following three $\Ye$ ranges:
high $\Ye$ ($\Ye = 0.30-0.40$, no lanthanide),
intermediate $\Ye$ ($\Ye = 0.20-0.30$,
lanthanide fraction of $\sim 5 \times 10^{-3}$),
and low $\Ye$ ($\Ye = 0.10-0.20$, lanthanide fraction of $\sim 0.1$).
The heating rate and abundances are averaged over the $\Ye$ range
above by using single-$\Ye$ nucleosynthesis calculations
with a step of $\Delta \Ye = 0.01$ by \citet{wanajo14}.
The thermalization efficiencies of $\gamma$-rays, $\alpha$ particles,
$\beta$ particles, and fission are separately taken into account
by analytically estimating characteristic timescales \citep{barnes16}.

  Figure \ref{fig:kappa_time} shows the time evolution of Planck mean opacity
  at $v=0.1c$.
  It shows the total opacity but the bound-bound opacity is always dominant.
  From $t=1$ to 10 days, the opacity increases with time for
  low and intermediate $\Ye$ cases,
  while it slowly decreases for high $\Ye$ case.
  The time evolution of the expansion opacity (Equation (\ref{eq:kappa}))
  is controled by the competition between
  the term of $1/\rho t$, which increases with time as $t^{2}$,
  and the summation of $1-e^{-\tau}$, which generally decreases with time
  as the lines get weaker for as the density decreases.
  The decrease of the line strength is more significant in the high $\Ye$
  case, which results in the temporal decrease of the opacity.
  Note that even with the increase in the opacity
  (for the low and intermediate $\Ye$ cases),
  the optical depth of the ejecta, $\tau \sim \kappa \rho R$, decreases.  
  In addition to the overall trend, the opacities shows temporal variation
  reflecting the temperature evolution, as shown in Figure \ref{fig:opacity_ye}.
  A significant decrease at $t \sim 10$ days corresponds to the
  sharp drop of the opacities at low temperature $T < 2000$ K.  

Overall degree of time variation is about an order of magnitude
  from $t = 1$ to 10 days.
  This gives a caveat to the use of a constant opacity in the analysis
  of kilonova light curves, although it is useful to derive physical parameters.
  Figure \ref{fig:kappa_time} also shows a hypothetical case where
  the abundance of the high $\Ye$ model is replaced with Fe.
  As discussed in Section \ref{sec:mixture}, Fe is not a representative
  element for $d$-shell elements.
  Therefore, the use of Fe for high $\Ye$ opaicty
  underestimates the opacity
  by a factor of 2-5 up to $t=2-3$ days and
  by a factor of more than 10 at later time.

\subsection{Light curves and spectra}
\label{sec:lc}

Finally we show the emergent light curves and spectra
  from the simulations in the previous section.
Compared with our previous calculations \citep{tanaka18}
using only Se ($p$-shell), Ru ($d$-shell), Te ($p$-shell),
Nd ($f$-shell), and Er ($f$-shell) as representative elements,
the light curves with new opacity data are more smooth
both in time and wavelength.
In particular, the use of representative elements can often exaggerate
emission in certain wavelengths.
At later time ($t \gsim 10$ days), only transitions from low-lying energy
levels contribute the opacities.
And thus, the use of small number of elements artificially
enhances contributions from transitions of these elements.
These effects are smeared out by properly including all the elements,
which results in smooth spectra.

As expected from the properties of the opacities,
the bolometric light curve of the model with high $\Ye$ has a shorter timescale
while that with low $\Ye$ has a longer timescale (Figure \ref{fig:lbol}).
Compared with the observed luminosity of AT2017gfo associated with
GW170817, the early ($t \sim 1-2$ days) light curve are most similar to the
high $\Ye$ model while the later light curve are most similar to
the intermediate $\Ye$ model.

These models also give a reasonable agreement with the multi-color
light curves of AT2017gfo (Figure \ref{fig:lc}),
although our models are very simple, and ejecta parameters such as mass and velocity are not tuned to reproduce the properties of AT2017gfo.
The high $\Ye$ model gives the early emission dominated in the optical wavelengths
while the intermediate $\Ye$ model gives the later emission dominated in
the NIR wavelengths.
It is emphasized that the optical/NIR flux ratio reflects
the wavelength dependence of the opacities, and thus,
cannot be accurately predicted by the calculations with a gray opacity.

The low $\Ye$ model overproduces the total luminosity and gives too red color,
which suggests that such a low $\Ye$ component with a lanthanide fraction of
$X({\rm La}) \sim 0.1$ is not dominant ($\Mej \ll 0.03 \Msun$).
This is consistent with a relatively low lanthanide fraction
$X({\rm La}) \lsim 0.01$ estimated by the spectral and light curve
modelling \citep{chornock17,kasen17,kilpatrick17,tanaka17,tanvir17}.

\begin{figure*}
  \begin{center}
      \includegraphics[scale=1.0]{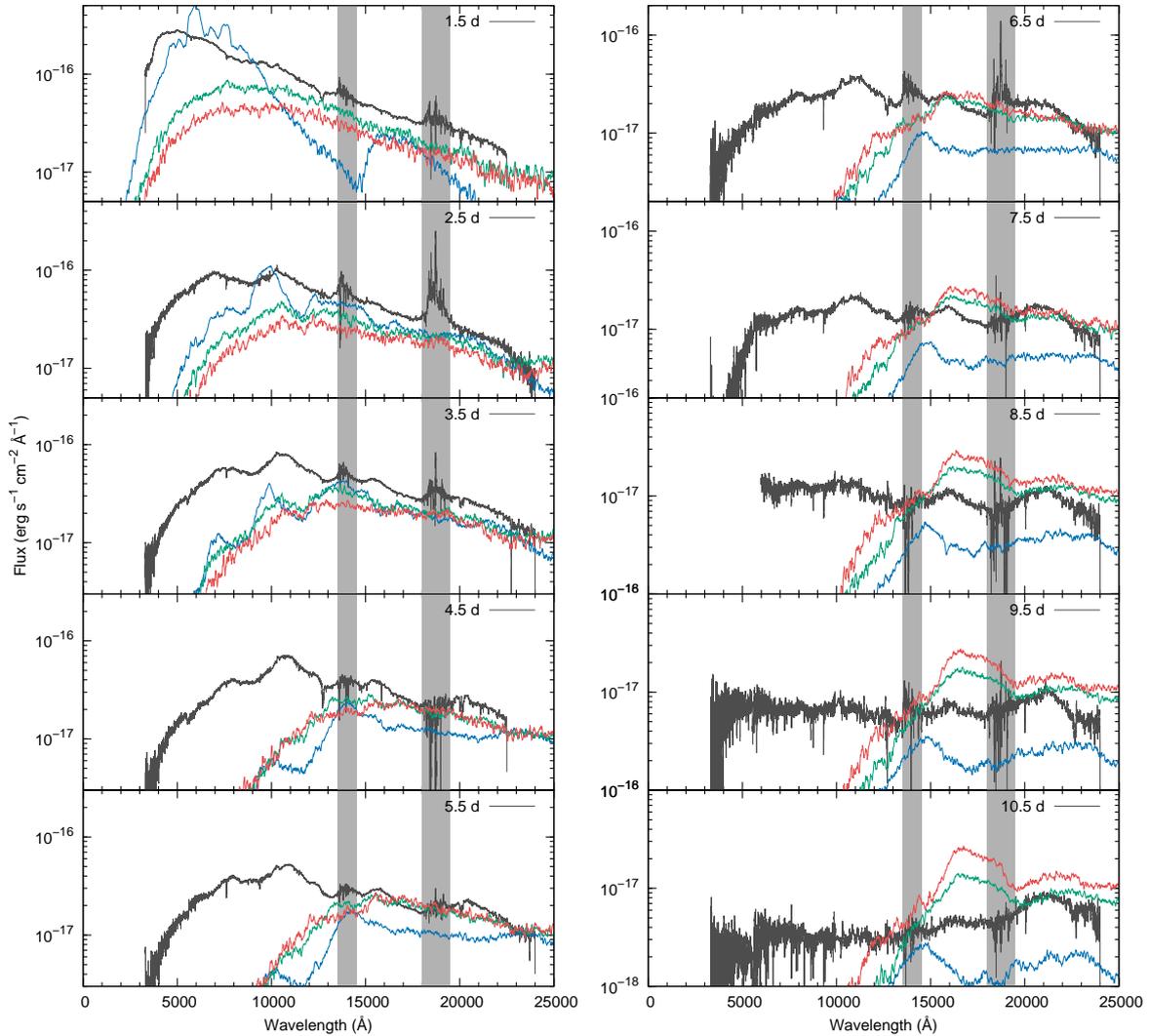} 
\caption{
  Spectral evolution of the models with
  high $\Ye$ ($\Ye = 0.30-0.40$, blue),
  intermediate $\Ye$ ($\Ye = 0.20-0.30$, green),
  and low $\Ye$  ($\Ye = 0.10-0.20$, red)
  compared with the spectra of GW170817/AT2017gfo
  taken with VLT/X-Shooter (\citealt{pian17}; \citealt{smartt17}, taken
  through the WISeREP, \citealt{yaron12}).
  The shaded areas show the wavelength ranges heavily affected by
  the atmospheric absorption.
\label{fig:spec}  
  }
\end{center}
\end{figure*}

The spectral features in our models are of interest
because this is the first systematic calculations
with the atomic data of the $r$-process elements.
Figure \ref{fig:spec} compares the model spectra
with the observed spectra of GW170817/AT2017gfo with VLT/X-Shooter
\citep{pian17,smartt17}.
The models capture overall spectral shape and its evolution:
the high $\Ye$ model gives a similar shape of the optical spectra
at early phases while intermediate $\Ye$ model gives a similar 
NIR flux level at later phases.

However, detailed spectral features are not necessarily
consistent between the observations and models.
This is not surprising because our atomic data do not have
an enough accuracy for each transition wavelength.
To identify the spectral features, we need to use
either well-calibrated (but not complete) atomic data
as done by \citet{tanaka13} and \citet{watson19}
or very accurate atomic calculations as done by \citet{gaigalas19}
and \citet{radziute20}.

There are two potentially important drawbacks in our models.
One is too narrow spectral features in the early spectra.
This is due to the assumption of $\langle v \rangle = 0.1c$ in our model.
The observed broader features indicate that
the line-forming region of the blue component should have $v > 0.1c$.
In fact, such high velocities of the blue component have been suggested
\citep[\eg][]{kilpatrick17,muccully17,nicholl17,shappee17}.
However, it was based on comparison with previous models,
which could exaggerate the spectral features by
the incompleteness in the atomic data.
The comparison with our new model with the complete opacity data
securely confirms the necessity of the high velocity for the blue component.

The other is the deficit of the optical flux at $t \gsim 5$
days after the merger.
It is difficult to keep the optical flux at $t \gsim 5$ days
because the optical flux in the high $\Ye$ model declines too quickly
and those in the intermediate and low $\Ye$ models are suppressed too much.
This difficulty remains even by changing ejecta mass and velocity.
We may obtain a better agreement by assuming a lanthanide fraction
  somewhat lower than that in the intermediate $\Ye$ model
  ($4.8 \times 10^{-2}$).
  Such a relatively small lanthanide fraction is also supported by
  the modelling of the light curves \citep{chornock17,kilpatrick17,tanaka17}.
  Although such a intermediate lanthanide fraction requires
  a fine tuning of $\Ye$,
  it can be naturally realized by the mixing in the ejecta 
  as pointed by \citet{metzger18}.
Note that the interplay between multiple ejecta components also influences
the light curve at this epoch \citep{kawaguchi18,kawaguchi20}.
Alternatively, this difficulty might point out
the necessity of more advanced radiative transfer calculations
by taking into account non-LTE or fluorescence of numerous transitions,
which are known to be important in supernovae
\citep[\eg][]{baron99,pinto00,mazzali00,dessart05}.

%%%%%%%%%%%%%%%%%%%%%%%%%%%%%%%%%%%%%%%%%%%%%%%%%%%% 
% Figure: Spectra
%%%%%%%%%%%%%%%%%%%%%%%%%%%%%%%%%%%%%%%%%%%%%%%%%%%% 
%%%%%%%%%%%%%%%%%%%%%%%%%%%%%%%%%%%%%%%%%%%%%%%%%%%%

%%%%%%%%%%%%%%%%%%%%%%%%%%%%%%%%%%%%%%%%%%%%%%%%%%%% 
% Section: Summary
%%%%%%%%%%%%%%%%%%%%%%%%%%%%%%%%%%%%%%%%%%%%%%%%%%%% 

\section{Summary}
\label{sec:summary}

We perform the first systematic atomic structure calculations for
neutral atoms and singly, doubly, and triply ionized ions of
the elements from Fe ($Z=26$) to Ra ($Z=88$)
to understand the elemental variation of the bound-bound
  opacities in the NS merger ejecta.
We find that the distributions of energy levels
tend to be shifted to higher energy for increasing number of electrons
in each shell.
Also, the total number of excited levels is the highest for the half-closed,
most complex elements.
The combination of these two effects determines 
degree of contributions to the opacities.
For typical temperature of kilonova ($T \sim 5,000$ K),
elements with lower number of electrons have
bigger contributions to the opacity thanks to the
relatively low-lying energy levels.
By this reason, Fe is not a good representative for the opacity of
lanthanide-free ejecta.
For a higher temperature ($T \gsim 10,000$ K),
elements with more electrons start to contribute
because more transitions from excited levels become active.

The average opacities of mixture of $r$-process elements are
$\kappa \sim 20-30 \ {\rm cm^2 \ g^{-1}}$ for $\Ye \le 0.20$,
$\kappa \sim 3-5  \ {\rm cm^2 \ g^{-1}}$ for  $\Ye = 0.25-0.35$,
and $\kappa \sim 1  \ {\rm cm^2 \ g^{-1}}$ for  $\Ye = 0.40$
at $T = 5,000-10,000$ K ($\rho = 1 \times 10^{-13} \ {\rm g \ cm^{-3}}$
and $t=$ 1 day).
Radiative transfer simulations with the new opacity data
  show that, even with the same abundance,
  the opacity in the ejecta changes with time.
  The opacity decreases with time
  for the model with high $\Ye$ ($\Ye = 0.30-0.40$, no lanthanide),
  while it increases and then decreases for the models with
  intermediate $\Ye$ ($\Ye = 0.20-0.30$, lanthanide fraction of $\sim 5 \times 10^{-3}$) and low $\Ye$ ($\Ye = 0.10-0.20$, lanthanide fraction of $\sim 0.1$).
  Overall variation is about an order of magnitude from $t=1$ to 10 days.

We confirm that multi-component ejecta are necessary to reproduce
the observed properties of GW170817/AT2017gfo.
The early blue part is best explained by the high $\Ye$ model
while the late NIR part is more similar to
the model with intermediate $\Ye$.
The model with low $\Ye$ overproduces the NIR light curves,
which suggests that such a low $\Ye$ component is not dominant
($\Mej \ll 0.03 \Msun$).

Although our calculations provide opacities of a wide range of
$r$-process elements,
the detailed spectral features in the model cannot be 
compared with the observed spectra because
our atomic data only focus on statistical properties
and do not have enough accuracies in the transition wavelengths.
To identify spectral features, 
combined use of accurate, well-calibrated (though not complete) atomic data
will be important.

\section*{Acknowledgements}

We thank Shinya Wanajo for providing the results of
nucleosynthesis calculations,
Michel Busquet for the generous support on the HULLAC code,
and Brian Metzger for giving fruitful suggestions.
MT and KK thank the Yukawa Institute for Theoretical Physics
for support in the framework of
International Molecule-type Workshop (YITP-T-18-06),
where a part of this work has been done.
Numerical simulations presented in this paper 
were carried out with Cray XC30 and XC50
at Center for Computational Astrophysics,
National Astronomical Observatory of Japan.

This research was supported by JSPS Bilateral Joint Research Project,
Inoue Science Research Award from Inoue Foundation for Science,
the Grant-in-Aid for Scientific Research from
JSPS (16H02183,19H00694,20H00158) and MEXT (17H06363),
the NINS program of Promoting Research by Networking
among Institutions (Grant Number 01411702).
GG thanks the Research Council of
Lithuania for funding his research (grant No. S-LJB-18-1).

%%%%%%%%%%%%%%%%%%%%%%%%%%%%%%%%%%%%%%%%%%%%%%%%%%



\bibliographystyle{mnras}
%\bibliography{../../Reference/reference} % if your bibtex file is called example.bib

%%%%%%%%%%%%%%%%%%%%%%%%%%%%%%%%%%%%%%%%%%%%%%%%%%

%%%%%%%%%%%%%%%%% APPENDICES %%%%%%%%%%%%%%%%%%%%%

\appendix

\section{Configurations used in the atomic calculations}

%%%%%%%%%%%%%%%%%%%%%%%%%%%%%%%%%%%%%%%%%%%%%%%%%%%% 
% Figure: Elevel example II
%%%%%%%%%%%%%%%%%%%%%%%%%%%%%%%%%%%%%%%%%%%%%%%%%%%% 
\begin{figure*}
  \begin{center}
    \begin{tabular}{cc}
    \includegraphics[scale=0.9]{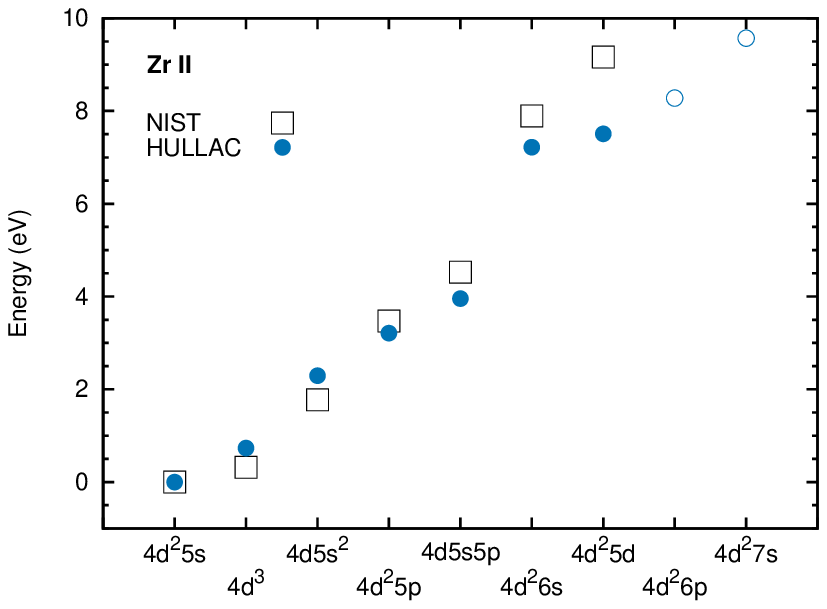} &
    \includegraphics[scale=0.9]{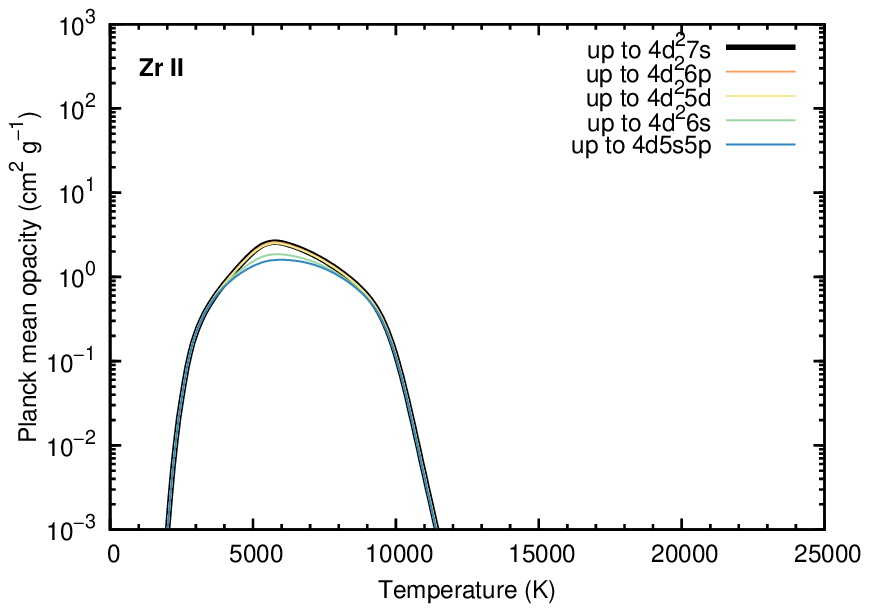}  \\
    \includegraphics[scale=0.9]{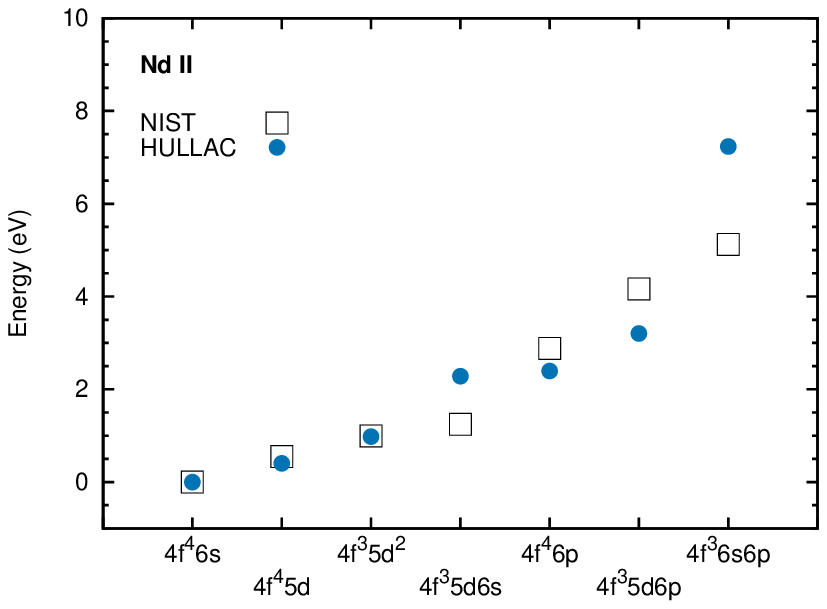} &
    \includegraphics[scale=0.9]{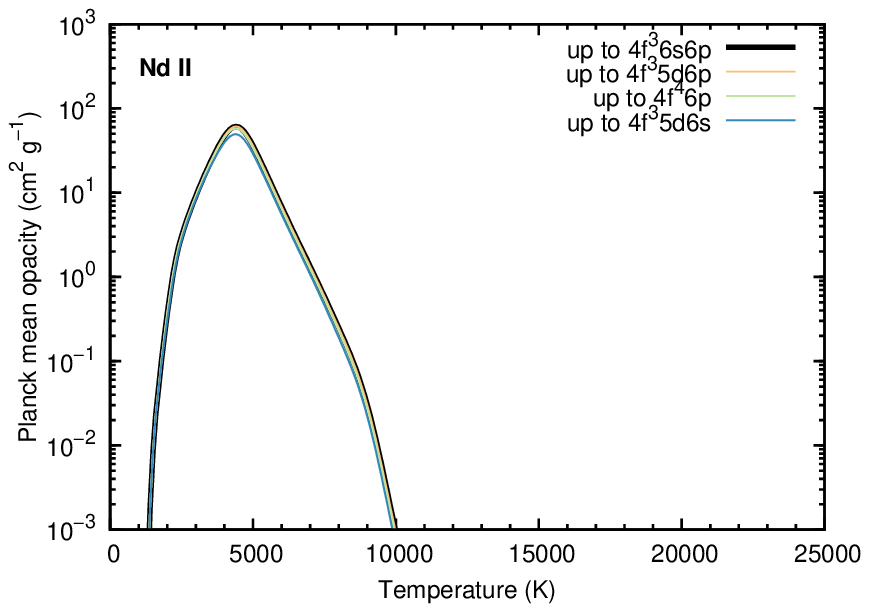}  \\
    \includegraphics[scale=0.9]{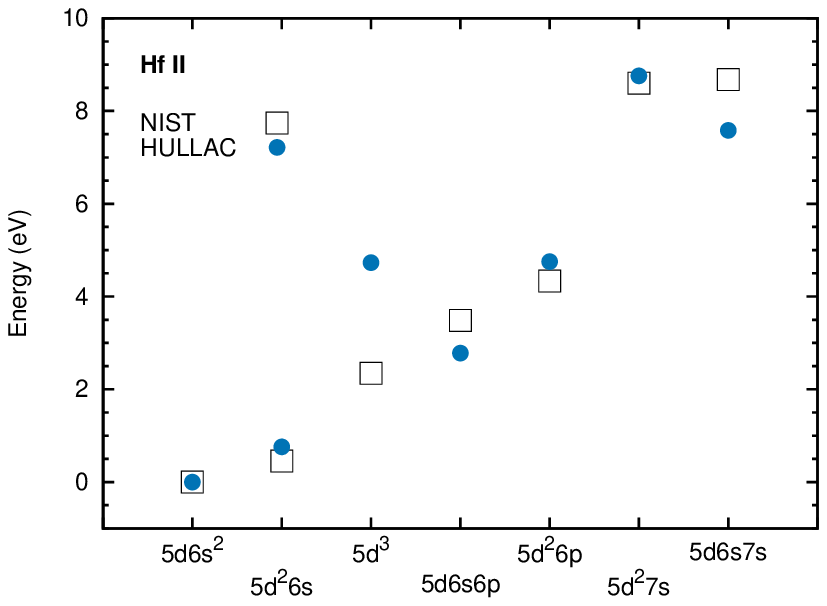} &
    \includegraphics[scale=0.9]{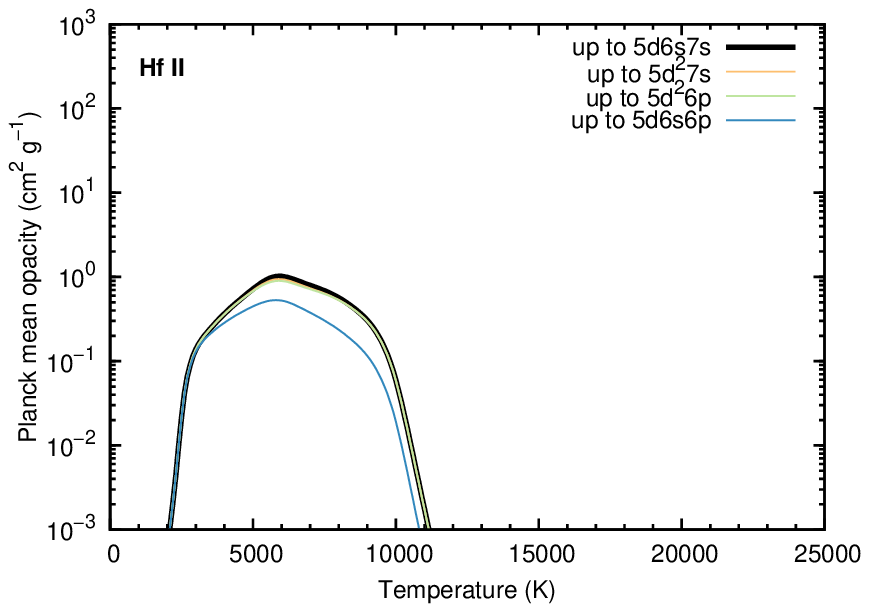}  \\    
    \end{tabular}
\caption{
  \label{fig:exampleII}
  Left: Lowest energy level of each configuration for Zr II ($Z=40, d$), Nd II ($Z=60, f$) , and Hf II ($Z=72, f$). Blue circles show our calculations while black squares show the data in the NIST ASD \citep{kramida18}.
  For our calculations, filled circles represent the configurations included in our default set and open circles represent the configurations added for the convergence studies of the opacities.
  Right: Planck mean opacities as a function of temperature, which are calculated with limited sets of configurations.
Our default calculations include the configurations up to $4d^25d$ (Zr II), $4f^36s6p$ (Nd II), and $5d6s7s$ (Hf II).
}
\end{center}
\end{figure*}
%%%%%%%%%%%%%%%%%%%%%%%%%%%%%%%%%%%%%%%%%%%%%%%%%%%%

%%%%%%%%%%%%%%%%%%%%%%%%%%%%%%%%%%%%%%%%%%%%%%%%%%%% 
% Figure: Elevel example II
%%%%%%%%%%%%%%%%%%%%%%%%%%%%%%%%%%%%%%%%%%%%%%%%%%%% 
\begin{figure*}
  \begin{center}
    \begin{tabular}{cc}
    \includegraphics[scale=0.9]{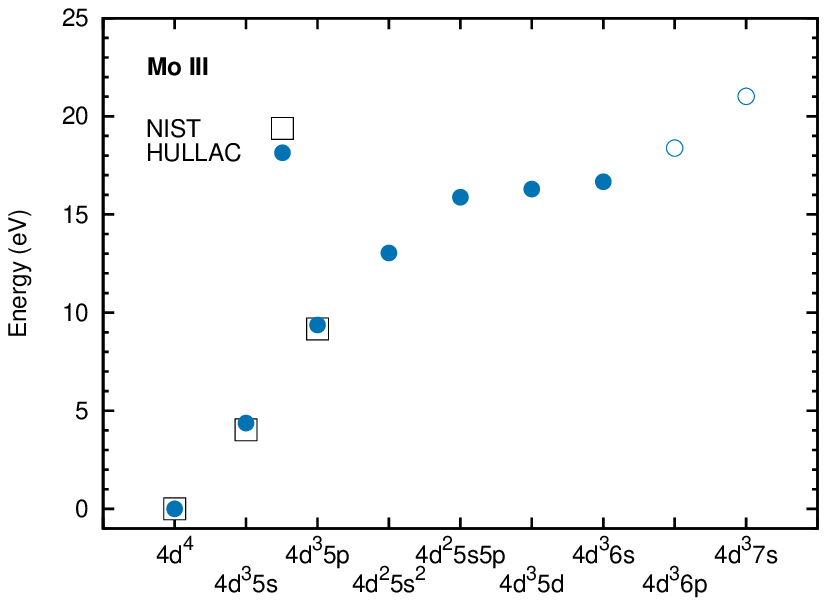} &
    \includegraphics[scale=0.9]{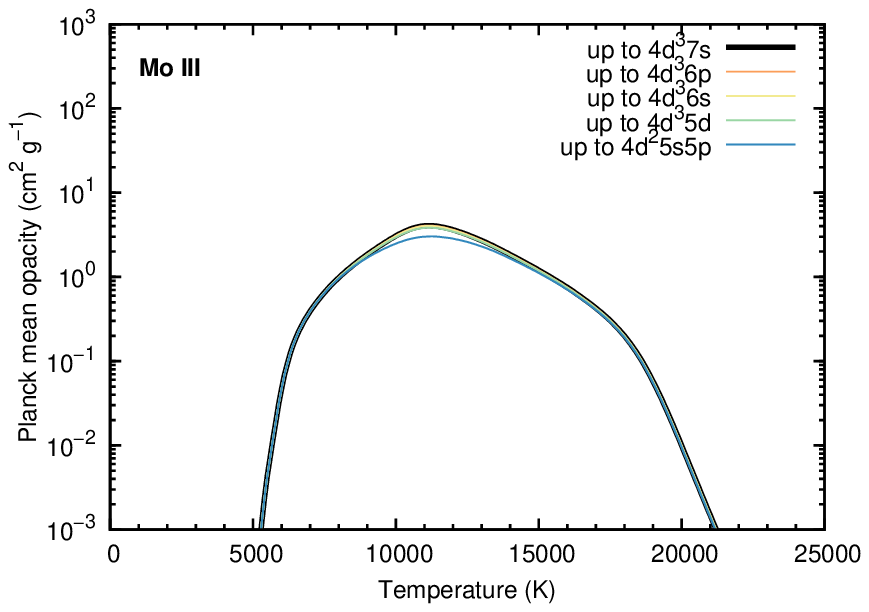}  \\
    \includegraphics[scale=0.9]{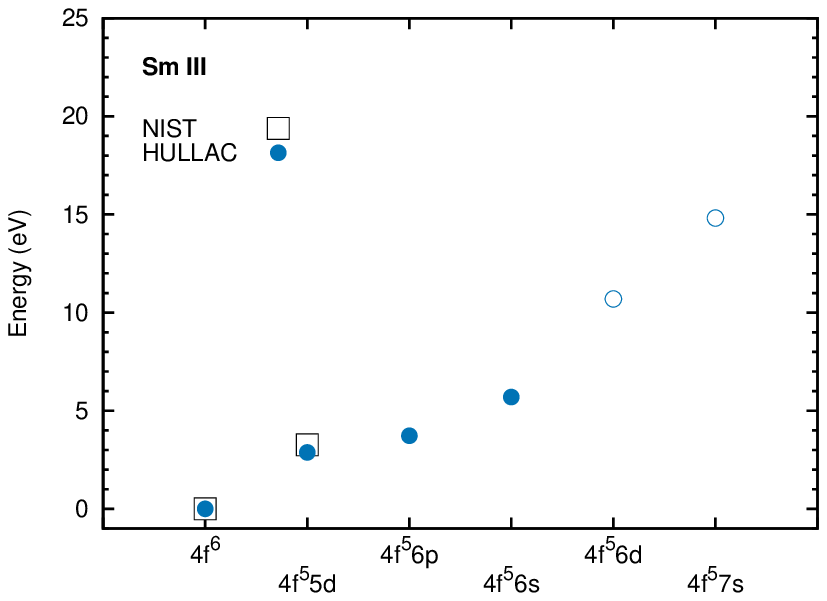} &
    \includegraphics[scale=0.9]{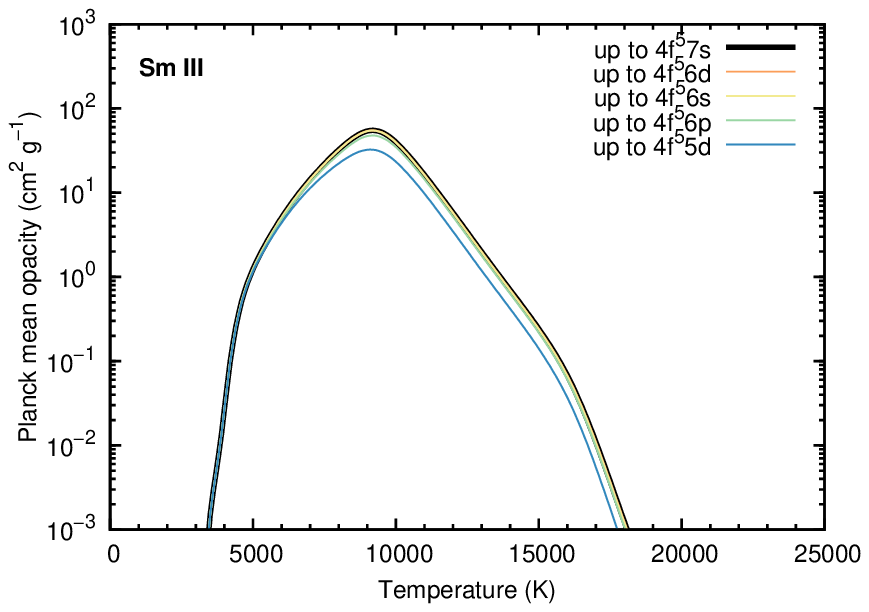}  \\
    \includegraphics[scale=0.9]{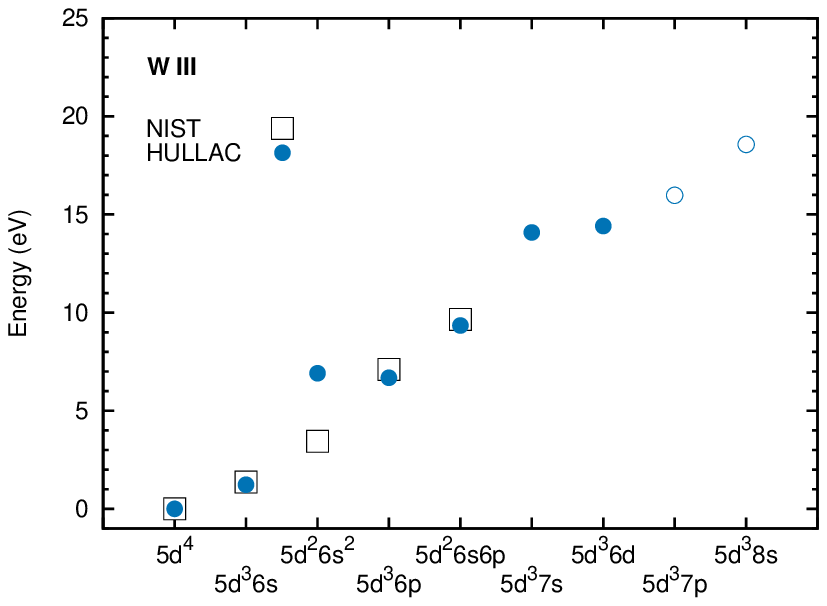} &
    \includegraphics[scale=0.9]{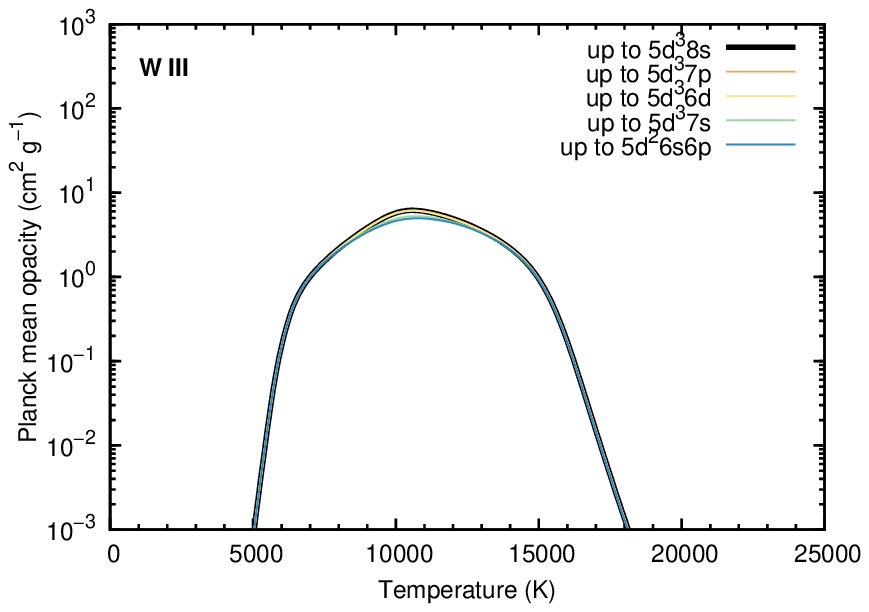}  \\    
    \end{tabular}
\caption{
  \label{fig:exampleIII}
  Same as Figure \ref{fig:exampleII} but for 
  Mo III ($Z=42, d$), Sm III ($Z=62, f$), and W III ($Z=74, d$).
Our default calculations include the configurations up to $4d^36s$ (Mo III), $4f^56s$ (Sm III), and $5d^36d$ (W III).
}
\end{center}
\end{figure*}
%%%%%%%%%%%%%%%%%%%%%%%%%%%%%%%%%%%%%%%%%%%%%%%%%%%%

Table 2 summarizes our atomic calculations.
For each ion, included configurations, the total number of levels,
the total number of transitions,
and the total number of transitions whose higher energy levels
are below the ionization threshold
(treated as bound-bound transitions in this paper) are given.
All the configurations given in the table are taken into account
in the RCI calculations.
The configurations used for the energy minimization are shown in bold font.

In Figures \ref{fig:exampleII}  and \ref{fig:exampleIII},
typical accuracy of our atomic calculations is shown
for the selected elements: Zr II ($Z=40, d$), Nd II ($Z=60, f$) , and Hf II ($Z=72, f$) which are important opacity source around $T=5,000$ K
and Mo III ($Z=42, d$), Sm III ($Z=62, f$), and W III ($Z=74, d$)
which are important around $T=10,000$ K.
In the figures, the lowest energy level for each configuration
is compared with that in the NIST ASD \citep{kramida18}.
As also shown in Figure \ref{fig:deltae}, a typical accuracy is
about $<20$ \% for singly ionized ions.
The number of available data is smaller for more ionized ions
(see top panel of Figure \ref{fig:deltae}).

Impacts of the included configurations to the opacities are
shown in the right panels of Figures \ref{fig:exampleII}
and \ref{fig:exampleIII}.
Different lines present the Planck mean opacities calculated with
limited sets of configurations.
For each ion, the atomic model is kept the same,
and energy levels of certain configurations are removed
to see the impact to the opacity.
With our default choice of configuration
(shown in filled blue circles in the left panels),
  the Planck mean opacities converge within 10\%.
When more configurations are included, the number of transitions does increase.
However, because these transitions have a large energy differences
(short wavelengths) or they are from highly excited levels,
they do not largely contribute to the overall opacities.

%%%%%%%%%%%%%%%%%%%%%%%%%%%%%%%%%%%%%%%%%%%%%%%%%%%% 
% Figure: Sobolev approximation
%%%%%%%%%%%%%%%%%%%%%%%%%%%%%%%%%%%%%%%%%%%%%%%%%%%% 
\begin{figure}
  \begin{center}
    \includegraphics[scale=0.9]{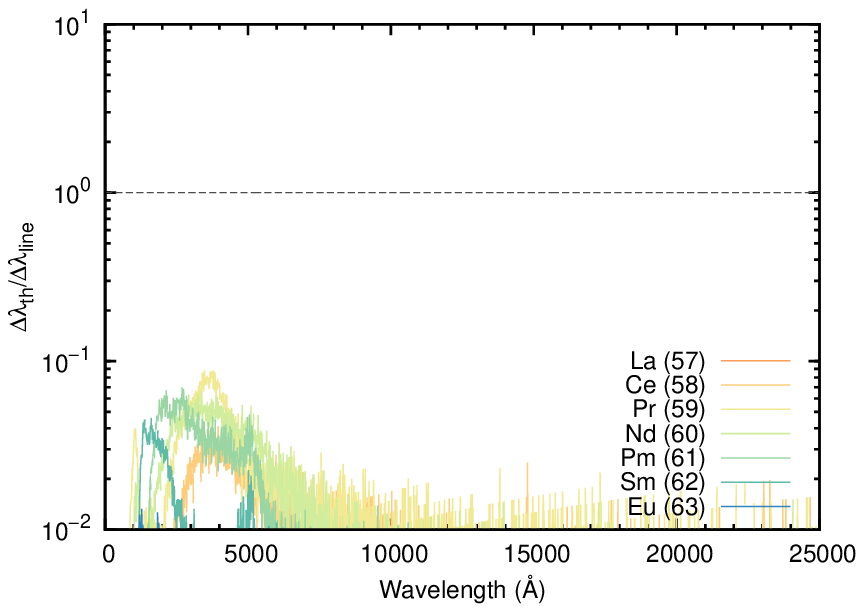} \\
    \includegraphics[scale=0.9]{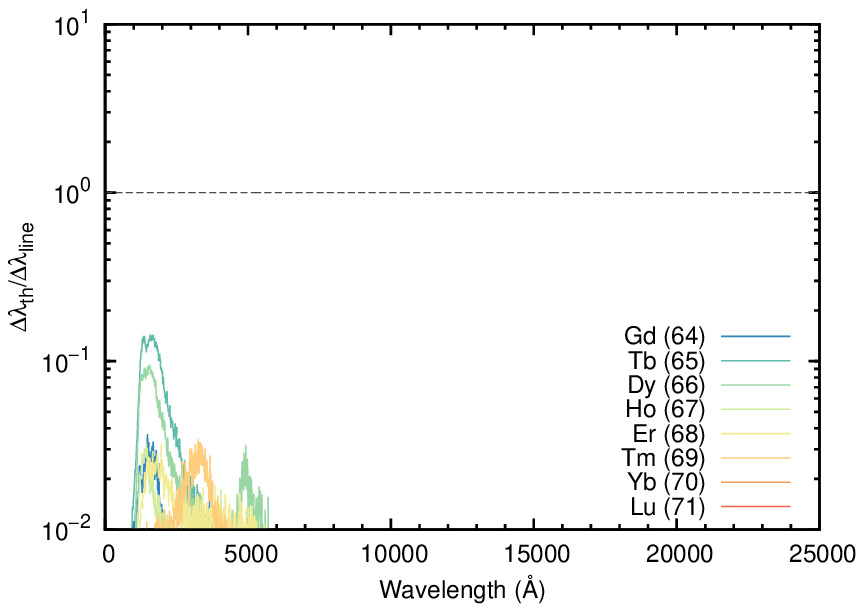} \\
    \includegraphics[scale=0.9]{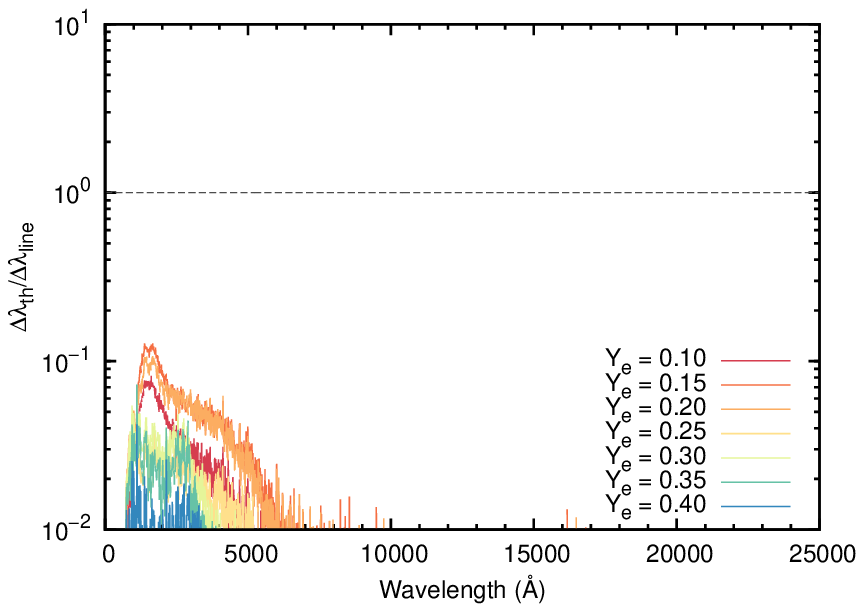} \\    
\caption{
  \label{fig:ratio}
  The ratio of the thermal width of the line $\Delta \lambda_{\rm th}$ to
  the typical wavelength spacing of the transitions $\Delta \lambda_{\rm line}$
  for the ejecta with $\rho = 1 \times 10^{-13} \ {\rm g \ cm^{-3}}$ and
  $T = 5000 $ K at $t=$ 1 day after the merger.
  The line spacing is evaluated for the strong lines with $\tau_l > 1$.
  Top and middle panels show the ratio for single-element ejecta (lanthanide)
  and the bottom panel shows the same for the mixture of the elements.
  For the mixture of the elements,
  the thermal width is calculated with the averaged mass number.
  Below the dashed line, the condition of
  $\Delta \lambda_{\rm line} > \Delta \lambda_{\rm th}$ is satisfied.
}
\end{center}
\end{figure}
%%%%%%%%%%%%%%%%%%%%%%%%%%%%%%%%%%%%%%%%%%%%%%%%%%%%

\section{Validity of the Sobolev approximation}

  We discuss the validity of the Sobolev approximation,
  which we use to evaluate the bound-bound opacities in the NS merger ejecta.
  When the wavelength spacing of the lines ($\Delta \lambda_{\rm line}$)
  becomes as small as the thermal width of the lines ($\Delta \lambda_{\rm th}$),
  overlap of the lines becomes severe and the Sobolev approximation is not applicable \citep{kasen13}.
  In a typical condition of the ejecta at $t=1$ day, the thermal velocity is
  $v_{\rm th} \simeq 0.7 \ {\rm km \ s^{-1}} (A/150)^{-1/2} (T/5,000 \ {\rm K})^{1/2}$,
  where $A$ is the mass number.
  Therefore, the thermal width of the line is
  $\Delta \lambda_{\rm th} = (v_{\rm th}/c)  \lambda \simeq 0.01$ \AA\ $(A/150)^{-1/2} (T/5,000 \ {\rm K})^{1/2} (\lambda / 5,000$ \AA).
  We can calculate typical line spacing by $\Delta \lambda_{\rm line} = \Delta \lambda / N$,
  where $\Delta \lambda$ is the wavelength bin in the calculation and $N$ is the number of
  the strong lines in the bin.
  Figure \ref{fig:ratio} shows the ratio $\Delta \lambda_{\rm th}/\Delta \lambda_{\rm line}$
  for the ejecta with pure lanthanide composition (top and middle panels)
  and for the ejecta with the mixture of the elements (bottom panels).
  The line spacing is evaluated by counting the number of lines with $\tau_l > 1$.
  For all the cases, the ratio $\Delta \lambda_{\rm th}/\Delta \lambda_{\rm line}$
  is smaller than unity, i.e., $ \Delta \lambda_{\rm line} > \Delta \lambda_{\rm th}$,
  at the relevant wavelength range.
  This relation holds even if the line spacing is evaluated by including weaker lines with $\tau > 0.1$.
  Therefore, the use of Sobolev optical depth is a sound approximation
  even in the lanthanide-rich ejecta of NS merger.

%%%%%%%%%%%%%%%%%%%%%%%%%%%%%%%%%%%%%%%%%%%%%%%%%%%% 
% Table: Atomic calculations %%%%%%%%%%%%%%%%%%%%%%%
%%%%%%%%%%%%%%%%%%%%%%%%%%%%%%%%%%%%%%%%%%%%%%%%%%%% 

% 1 %%%%%%%%%%%%%%%%%%%%%%%%%%%%%%%%%%%%%%%%%%%%%%%%
\begin{table*}
  \centering
  \caption{Summary of HULLAC calculations. The last column shows the number of transitions whose upper level is below the ionization potential.}
  \label{tab:hullac}

\end{table*}
%%%%%%%%%%%%%%%%%%%%%%%%%%%%%%%%%%%%%%%%%%%%%%%%%%%% 

% 2 %%%%%%%%%%%%%%%%%%%%%%%%%%%%%%%%%%%%%%%%%%%%%%%%
\begin{table*}
  \centering
  \contcaption{Summary of HULLAC calculations. The last column shows the number of transitions whose upper level is below the ionization potential.}
  %
\end{table*}
%%%%%%%%%%%%%%%%%%%%%%%%%%%%%%%%%%%%%%%%%%%%%%%%%%%% 

% 3 %%%%%%%%%%%%%%%%%%%%%%%%%%%%%%%%%%%%%%%%%%%%%%%%
\begin{table*}
  \centering
  \caption{Summary of HULLAC calculations. The last column shows the number of transitions whose upper level is below the ionization potential.}
  %
\end{table*}
%%%%%%%%%%%%%%%%%%%%%%%%%%%%%%%%%%%%%%%%%%%%%%%%%%%% 

% 4 %%%%%%%%%%%%%%%%%%%%%%%%%%%%%%%%%%%%%%%%%%%%%%%%
\begin{table*}
  \centering
  \contcaption{Summary of HULLAC calculations. The last column shows the number of transitions whose upper level is below the ionization potential.}
  %
\end{table*}
%%%%%%%%%%%%%%%%%%%%%%%%%%%%%%%%%%%%%%%%%%%%%%%%%%%% 

% 5 %%%%%%%%%%%%%%%%%%%%%%%%%%%%%%%%%%%%%%%%%%%%%%%%
\begin{table*}
  \centering
  \caption{Summary of HULLAC calculations. The last column shows the number of transitions whose upper level is below the ionization potential.}
  %
\end{table*}
%%%%%%%%%%%%%%%%%%%%%%%%%%%%%%%%%%%%%%%%%%%%%%%%%%%% 

% 6 %%%%%%%%%%%%%%%%%%%%%%%%%%%%%%%%%%%%%%%%%%%%%%%%
\begin{table*}
  \centering
  \contcaption{Summary of HULLAC calculations. The last column shows the number of transitions whose upper level is below the ionization potential.}
  %
\end{table*}
%%%%%%%%%%%%%%%%%%%%%%%%%%%%%%%%%%%%%%%%%%%%%%%%%%%% 

% 7 %%%%%%%%%%%%%%%%%%%%%%%%%%%%%%%%%%%%%%%%%%%%%%%%
\begin{table*}
  \centering
  \caption{Summary of HULLAC calculations. The last column shows the number of transitions whose upper level is below the ionization potential.}
  %
 & 60 & 574 & 457\\ 

    \hline
  \end{tabular}
\end{table*}
%%%%%%%%%%%%%%%%%%%%%%%%%%%%%%%%%%%%%%%%%%%%%%%%%%%% 

% Don't change these lines
\bsp	% typesetting comment
\label{lastpage}
\end{document}